\documentclass[fleqn,usenatbib,useAMS]{mnras}

\newif\ifshort
\shorttrue % if using short figs
\usepackage[dvips]{graphics}
\usepackage{graphicx}
\usepackage{amsmath}
\usepackage{epsfig}
\usepackage[latin1]{inputenc}
\usepackage{hyperref}

\usepackage[T1]{fontenc}
\usepackage{ae,aecompl,url}
\usepackage{multicol,placeins}
\usepackage{cuted,mathtools}
\usepackage{dcolumn,caption,booktabs}

\title[Orbit Integration of Young Stellar Associations]{Birth Sites of Young Stellar Associations and Recent Star Formation in a Flocculent Corrugated Disk}
\author[Quillen et al.]{
Alice C. Quillen$^1$\thanks{E-mail: alice.quillen@rochester.edu},  
Alex R. Pettitt$^2$,  %alex@astro1.sci.hokudai.ac.jp
Sukanya Chakrabarti$^3$, %chakrabarti@astro.rit.edu
Yifan Zhang$^1$, %yzh250@u.rochester.edu 
\newauthor
Jonathan Gagn\'e$^{4,5}$, % [0000-0002-2592-9612] jonathan.gagne.1@gmail.com 
and 
Ivan Minchev$^6$\\ % ivan.minchev@aip.de  iminchev1@gmail.com
$^1$Department of Physics and Astronomy, University of Rochester, Rochester, NY 14627, USA\\
$^2$Department of Physics, Faculty of Science, Hokkaido University, Sapporo 060-0810, Japan\\
$^3$School of Physics and Astronomy, Rochester Institute of Technology, 84 Lomb Memorial Drive, Rochester, NY 14623, USA\\
$^4$Institut de Recherche sur les Exoplan\`etes, Universit\'e de Montr\'eal, Pavillon Roger-Gaudry, \\
\ \ \ \ \ \ \ \ \ PO Box 6128 Centre-Ville STN, Montr\'eal QC H3C 3J7, Canada\\
$^5$Plan\'etarium Rio Tinto Alcan, Espace pour la Vie, 4801 av. Pierre-de Coubertin, Montr\'eal, Qu\'ebec, Canada\\
$^6$Leibniz Institut f\"ur Astrophysik Potsdam (AIP), An der Sternwarte 16, D-14482, Potsdam, Germany\\
}
\hypersetup{draft}

\begin{document}
\maketitle
\begin{abstract}
With backwards orbit integration we estimate birth locations of young stellar associations and moving groups identified in the solar neighborhood that are younger than 70 Myr. The birth locations of most of these stellar associations are at smaller galactocentric radius than the Sun, implying that their stars moved radially outwards after birth.  Exceptions to this rule are the Argus and Octans associations, which formed outside the Sun's galactocentric radius.  Variations in birth heights of the stellar associations suggest that they were born in a corrugated  disk of molecular clouds, similar to that inferred from the current filamentary molecular cloud distribution and dust extinction maps. Multiple spiral arm features with different but near corotation pattern speeds and at different heights could account for the stellar association birth sites. We find that the young stellar associations are located in between peaks in the radial/tangential (UV) stellar velocity distribution for stars in the solar neighborhood. This would be expected if they were born in a spiral arm which perturbs stellar orbits that  cross it.  In contrast, stellar associations seem to be located near peaks in the vertical phase-space distribution, suggesting that the gas in which stellar associations are born moves vertically together with the low velocity dispersion disk stars.
\end{abstract}

%continuation of the local spur seen in masers or the 'Split' featured seen in extinction mapslabelled 
% by  \citet{lallement19} and  
%\begin{keywords}
%minor planets, asteroids, general \\ % < Planetary Systems >, 
%planets and satellites: dynamical evolution and stability % < Planetary Systems
%\end{keywords}

\section{Introduction}

Coeval groups of stars are tracers of the past sites of star formation in the Milky Way disk (e.g., \citealt{zuckerman04}).
As it orbits the Galaxy, a recently formed cluster, group or association of stars, retains information on its birth location.
However, birth locations are more uncertain for  older
clusters, due to uncertainties in the cluster age and in the galactic potential in which orbits
are integrated (e.g. \citealt{dias19}).  
Close encounters with star clusters and molecular clouds, cluster evaporation, 
and non-axisymmetric and time dependent gravitational forces associated with spiral arms, the Galactic bar,
passage through the Galactic plane,  
and external tidal forces on the Galaxy disk perturb stellar orbits and increase the error
in the estimated birth sites of stellar associations and clusters (e.g., \citealt{kraus20}). 

Because open clusters are expected to have predominantly been born in spiral arms,
their orbits and ages have been used to probe the nature of spiral structure \citep{dias05,quillen18a,dias19}. 
The Lin-Shu  hypothesis  \citep{linshu66,shu16}  posits that spiral structure is caused by  
 density waves moving through a galactic disk.  The speed of
the wave is described with an angular rotation rate called the pattern speed.  
In the `modal' view of spiral structure, a single multi-armed wave with a constant
pattern speed dominates  \citep{linshu66,bertin89,shu16}.
A direct method for measuring the pattern speed of spiral arms relies on estimating
the birth places of open clusters by integrating their orbits backwards (e.g., \citealt{dias05,naoz07,dias19}).
However, this method can give spurious results if spiral arms are not steady-state.  
Spiral arms could be transient  \citep{toomre81,sellwood84,wada11,grand12,wada13,baba16}, exhibiting  variations in their amplitude, 
pitch angle and pattern speed.  Additional gravitational perturbations arise from the time-dependent and flocculent nature of the gas response (and hence that of the young stars) even when there is a time-steady spiral structure in the old stars \citep{chakrabarti03}.  Multiple patterns could simultaneously be present 
 (e.g., \citealt{naoz07}) and interfere  (e.g., \citealt{quillen11,comparetta12}). 
External tidal perturbations on the Galaxy can induce spiral structure  (e.g., \citealt{quillen09,chakrabarti09,dobbs10,delavega15,pettitt17}).
The recent study by \citet{dias19} mitigates the errors inherent in backwards orbit
integration and from assuming steady state spiral structure by focusing on nearby (distances  less
than 5 kpc from the Sun) and younger than 50 Myr old open clusters
with accurately measured distances, space motions and ages.   

Open clusters are groups of coeval stars, with most stars 
remaining in a gravitationally bound clump since birth.
While they are not gravitationally bound entities, a number of moving groups and stellar associations
seen in the solar neighborhood  are also comprised of coeval groups of stars 
(e.g, \citealt{eggen83,zuckerman04,mamajek16,riedel17}).    
Young stellar associations are predominantly identified in the solar neighborhood (within about 150 pc of the Sun), 
and they span a range of ages,  including some younger than 50 Myr.  As most open clusters are more distant
than the known young stellar associations (the  open clusters studied by  \citealt{dias19} are within  a few kpc from the Sun),
 stellar associations 
give a complimentary view of recent star formation near the Sun.  We aim to probe the locations 
of spiral arm density peaks in the recent past by estimating the birth locations of coeval young stellar associations and moving groups.

Members of a young stellar association or young moving group in the solar neighborhood share similar space velocities, or 
velocity components $U,V,W$, in heliocentric polar Galactic coordinates, with typical velocity dispersions 
below a few km s$^{-1}$ (e.g., \citealt{binks15,mamajek16,riedel17,gagne18a}).
Young stellar associations are discovered by searching for nearby stars with 
similar proper motions and evidence of youth, with more detailed studies
 identifying additional members, confirming stellar association membership and finding substructures in the association 
 age, velocity and spatial distributions (e.g., \citealt{eggen83,delareza89,dezeeuw99,jaya00,mamajek99,binks15,pecaut16,mamajek16,gagne18a,gagne18b,gagne18c,gagne18d,meingast19,kos19,binks20,tian20}).

The brightest nearby young stars, are not randomly distributed  
in the Galaxy \citep{dezeeuw99,elias06,bouy15,zari18},
rather those within 150 pc of the Sun  appear to form a belt, known as the `Gould belt',
with an inclination of about $20^\circ$ with respect to the Milky Way mid-plane \citep{herschel,gould,perrot03}.
The Gould belt may be part of a larger vertical wavelike structure that is present in a filament of molecular cloud complexes 
\citep{zucker20} dubbed the `Radcliffe wave' \citep{alves20}.  Vertical corrugations in molecular gas filaments could be 
related to the spiral seen in the distribution of solar neighborhood stars in the vertical components  of 
phase-space $z, \dot z$  \citep{antoja18}.
The dynamics of the interstellar medium differs from that of stars.  Shocks and associated linear density
enhancements in the gas can be caused by the spiral structure, whereas phase wrapping after a tidal
perturbation (e.g., \citealt{minchev09,delavega15})
in the vertical components of phase-space \citep{candlish14,antoja18,bland19} occurs in the stars but not in the gas, and can persist in the phase-space structure of the stars for many crossing times.  In contrast, density disturbances in the gas disk dissipate after a dynamical time.  Prominent gas density perturbations in the Galaxy include the HI warp and planar disturbances \citep{Levine06warp, Levine06planar}.  The height above or below the Galactic plane of birth sites of young stars may reflect the past location of features like the Gould belt and the Radcliffe wave.  Vertical corrugations
in the molecular disk and the stellar phase-space spiral are suspected to have been excited by tidal perturbations on the outer Galaxy (e.g., \citealt{chakrabarti09,quillen09,purcell11,delavega15,antoja18,darling19}). 

%The belt contains
%three stream-like structures, named Scorpius-Canis Major, Vela, and Orion, (e.g.,  \citealt{})
%and is not only coherent in space but also characterized by sub-structures that have age sequences  
%(e.g., \citealt{pecaut16,wright18}).

Structure in the distribution of  VLBI observations of masers associated with high mass
star formation regions \citep{xu16,xu18,reid19}, resembles that seen in extinction maps 
\citep{rezaei18,green19,lallement19}, and in 
a 3D map of nearby molecular clouds that is based on  
combining stellar photometric data with stellar Gaia DR2 parallax measurements to infer the distances of nearby 
 dust clouds \citep{zucker20,alves20}.
Spiral features seen in extinction maps \citep{quillen02,rezaei18,green19,lallement19}, masers \citep{xu16,xu18,reid19} 
and molecular clouds \citep{zucker20} suggest that
within a few kpc of the Sun, the Milky Way contains multiple spiral arms with morphology more flocculent than grand design.
The recent open cluster study by \citet{dias19} matched open cluster birth locations to  Perseus, Local and Sagittarius arms as  traced by masers and HII regions and  following
logarithmic functions describing spiral arm peaks by \citet{reid14}.  The study of open cluster kinematics by
\citet{dias19}  found that these three spiral structures have pattern
speeds nearly corotating with the Sun.

With the recent advances in identifying and characterizing young stellar associations 
(e.g., \citealt{malo13,gagne18a,riedel17}),
  it is a good time to look for connections between them and spiral
arm candidates seen in the recently improved 
 molecular cloud and extinction maps.  To do this we integrate the orbits
of young stellar associations backwards to estimate their birth locations. 
We focus here on what we can learn from integrating the orbits of recent compilations
of young stellar associations and star formation regions (e.g., \citealt{gagne18a}).
Backwards orbit integration has been used to estimate kinematic ages of stellar associations
by  finding when their stellar members were likely to have been in proximity
\citep{delareza06,miretroig18,crundall19}.  Here we do not try to estimate kinematic ages, rather we search
for patterns in the history of recent star formation near the Sun.
We integrate orbits in three-dimensions to probe the relation between birth heights,
the corrugated molecular and  dust disk, and patterns seen in the stellar vertical phase-space
distribution \citep{antoja18,bland19,laporte19}.
%ages for the TW Hydrae Association \citep{delareza06} 

In section \ref{sec:fororbit} we specify coordinates, constants and the potential model needed
for orbit integration.
The sample of stellar associations, clusters and moving groups and their measurements
are described in section \ref{sec:sample}. %   Estimated birth locations for the stellar associations are listed in section \ref{sec:birth_loc}.

Results of backwards orbit integrations are discussed in section \ref{sec:results}.
In section \ref{sec:birth} we discuss estimated stellar association birth locations in two dimensions.
Birth heights above and below the plane are discussed in section \ref{sec:heights}.
In sections \ref{sec:rot} and \ref{sec:height_rot} we look at birth sites in rotating frames and discuss possible
molecular and extinction filaments that could be the current counterparts to
past sites of star formation.   In sections \ref{sec:uvw} and \ref{sec:zvz} we compare stellar association 
phase-space coordinates to the distributions of stars in the solar neighborhood.
A summary and discussion follows in section \ref{sec:sum}.

\section{Galactic potential model for orbit integration}
\label{sec:fororbit}

We specify the Galactic coordinate systems, and our notation for them in section \ref{sec:coord}.
We also review  
constraints on Galactic constants which are needed to carry out backwards orbit integrations.
The gravitational potential we use to integrate the orbits 
is described in section \ref{sec:pot}.

\subsection{Coordinate system and Galactic constants}
\label{sec:coord}

Numerous  works have used the transformations given by \citet{johnson87} to take stellar parallax, proper motion, position, and radial velocity measurements and compute  a heliocentric  coordinate ($x_h,y_h,z_h$) and velocity vector, $(U,V,W)$.
A heliocentric right-handed Cartesian coordinate system with origin at the Sun has Galactic coordinates 
\begin{equation}
(x_h,y_h,z_h) = d(\cos b \cos l, \cos b \sin l, \sin b),
\end{equation}
 where $b, l$ are Galactic latitude and longitude, respectively, and $d$ is the distance of the point from the Sun.
The positive $z_h$ axis is along the north Galactic pole.
Longitude $l=0$, latitude $b=0$ and small positive $x_h$ corresponds to a point
that is closer to the Galactic center than the Sun.
Galactic longitude $l = \pi/2$ and latitude $b=0$ gives positive $y_h$ axis pointing in the direction of Galactic rotation.
Heliocentric $U,V,W$ velocity components are in cylindrical coordinates with $U$ the radial component of velocity, $V$ the  tangential component and $W$ the vertical velocity component.
These velocity components are positive toward the Galactic center, in the direction
of Galactic rotation and in the direction of the North Galactic pole.
The velocities are in a heliocentric frame, so must be corrected for the solar motion
with respect to the local standard of rest (LSR).

The galactocentric Cartesian coordinate system $(x_g,y_g,z_g)$,  
has origin at the Galactic center.
We compute galactocentric Cartesian coordinates $(x_g,y_g,z_g)$  
from heliocentric $(x_h,y_h,z_h)$ Cartesian coordinates with
\begin{align}
(x_g, y_g, z_g) & = (-R_\odot  + x_h, y_h, z_h + z_\odot), 
\end{align}
where $z_{\odot}$ is the location of the Sun above or below the Galactic plane
and $R_\odot$ is the galactocentric radius of the Sun.   Based on trigonometric parallaxes of high-mass star forming regions  
\citet{reid19} find 
\begin{equation} z_\odot = 5.5 \pm 5.8\ {\rm pc} \end{equation} which agrees with that found by \citet{anderson19} based on positions of HII regions.  We adopt the value of $z_\odot=5.5$ pc in our orbit integrations.
However, we note that \citet{bennett19} find that 
$z_\odot= 20.8 \pm 0.3\ {\rm pc}$ from measurements of the vertical stellar distribution function. 
In this galactocentric coordinate system, Galactic rotation in the $x_g,y_g$ plane is clockwise about the origin. Galactocentric cylindrical coordinates $R_g, \theta_g, z_g$ can be computed using galactocentric azimuthal angle and radius
\begin{align}
\theta_g &= {\rm atan2}(y_g, x_g), \qquad
R_g = \sqrt{x_g^2 + y_g^2}.
\end{align}
The galactocentric azimuthal angle of the Sun, $\theta_\odot = \pi$ and the local standard of rest has clockwise rotation with 
angular rotation rate $\dot \theta_g <0$.  A point with radius $R_g = R_\odot$ 
and small and negative  $\theta - \theta_\odot$ 
 is in front of the local standard of rest in the direction of rotation. 

From heliocentric $U,V,W$ velocity components
we compute the galactocentric velocity in cylindrical  coordinates 
\begin{align}
(v_R, v_\theta, v_z) =\left(  -(U + U_\odot) , -(V + V_\odot + V_{LSR}), W + W_\odot \right) 
\end{align}
where $(U_\odot, V_\odot, W_\odot)$ is the peculiar velocity of the Sun in cylindrical coordinates with respect to
the local standard of rest (LSR).
For this velocity transformation we adopt a peculiar solar motion of 
\begin{equation}
  (U_\odot, V_\odot, W_\odot) =  (11.1^{+0.69}_{-0.75}, 12.24^{+0.47}_{-0.47}, 7.25^{+0.37}_{-0.36})\ {\rm km\ s}^{-1} 
\end{equation} 
based on an analysis of local stellar kinematics \citep{schonrich10}.

%The Gaia Collaboration \citep{katz18},  adopted local standard of rest (LSR) rotational velocity $V_{LSR} = 240\ {\rm km\ s}^{-1}$, and the Sun's galactocentric radius  $R_\odot = 8.34$ pc, following \citet{reid14}.

%However, with more recent analysis of trigonometric parallaxes of high mass star formation regions \citet{reid19} finds $V_{LSR} = 236 \pm 7\ {\rm km\ s}^{-1}$ and $R_\odot = 8.15 \pm 0.15\ {\rm kpc}$.

Using the pericenter passage of a star around the Galaxy's central supermassive black hole,
the \citet{gravity18} measures the galactocentric radius of the Sun
\begin{equation}
R_\odot = 8.122 \pm 0.031\ {\rm kpc}.
\end{equation}
This radius, the proper motion of the radio source associated with the Galaxy's central black hole, Sgr A*,  and the tangential component of the solar peculiar motion measured
by \citet{schonrich10} gives local standard of rest (LSR) rotational velocity 
\begin{equation}
V_{LSR} = 233 \pm 1.4\ {\rm km\ s}^{-1}.
\end{equation}
These values are consistent with those computed from  trigonometric parallaxes of high-mass star formation regions \citep{reid19} and we use these values in our orbit integrations.
These values give LSR angular rotation rate 
$\Omega_\odot = V_{LSR}/R_\odot = 28.7 \ {\rm km\ s}^{-1} {\rm kpc}^{-1} $.

\subsection{The Galactic potential and backwards orbit integration}
\label{sec:pot}

The stars in young stellar associations are  in nearly circular orbits that
remain  within 150 pc of  the Galactic plane.
Rather than use a mass model for the entire galaxy comprised of disk, bulge and halo
components (e.g., \citealt{robin03,deg19}), 
we use a local axisymmetric potential model that matches the slope of the
rotation curve near the Sun's galactocentric radius and measurements of the vertical acceleration of
stars above the Galactic plane.  A multiple mass component  Milky  Way  
model is not needed as we only integrate low eccentricity and low inclination orbits.
In cylindrical coordinates, we approximate the potential as
\begin{align}
 \Phi(R_g,z_g) = \Phi_R(R_g) + \Phi_z(z_g). \label{eqn:Phi}
\end{align}
As did \citet{darling19}, we adopt a static, 
axisymmetric potential function that is separable in the radial and vertical coordinates.
% {\color{red}[ARP: maybe a comment here about the justification of this separability?]}.  
%Our potential function is static, axisymmetric and separable 

We describe the slope of the Galactic rotation curve
with an exponent $\beta$ (following \citealt{dehnen99}, Appendix B).  
The tangential velocity of a star in a circular orbit in the Galactic plane 
\begin{align}
v_c(R_g) = % \sqrt{ R_g \frac{d \Phi_R }{dR_g} }=
 \left\{ \begin{array}{lll}
     V_{LSR}   & {\rm for} & \beta=0 \\
     V_{LSR} \left( \frac{R_g}{R_\odot} \right)^\beta  & {\rm for} & \beta \ne 0 
\end{array}
\right. \label{eqn:vc}
\end{align}
For this power law rotation curve, the radius of a circular orbit with $z$ component
of angular momentum $L$ is 
\begin{equation} 
 R_L(L) =  R_\odot \left( \frac{L}{R_\odot V_{LSR} }\right)^\frac{1}{1+\beta} \label{eqn:R_L}
 \end{equation}
 \citep{dehnen99}.
 The rotation curve (equation \ref{eqn:vc}) is consistent with radial potential function 
\begin{equation}
\Phi_R(R_g) =  \left\{ \begin{array}{lll}
	V_{LSR}^2 \ln \left( \frac{R_g}{R_\odot} \right)   & {\rm for} & \beta=0 \\
	\frac{V_{LSR}^2}{2 \beta} \left( \frac{R_g}{R_\odot} \right)^{2\beta}  & {\rm for} & \beta \ne 0. \\
\end{array}
\right. \label{eqn:Phi_R}
\end{equation}

With a mass model and terminal velocity constraints based on observations of atomic and molecular gas emission lines, 
\citet{mcgaugh19} estimates a rotation curve slope at the solar galactocentric radius $R_\odot$ of  
\begin{equation}
\left. \frac{dv_c(R_g)}{dR_g} \right|_{R_\odot}= -1.7 \pm 0.1\ {\rm km\ s}^{-1} \ {\rm kpc}^{-1} \label{eqn:slope1}.
\end{equation}
%over $9<R_g< 19$ kpc.  
This slope is the same as the one inferred from recent measurements of Oort's $A$ and $B$ constants
\citep{li19} and is  similar to  the slope 
\begin{equation}  \left. \frac{dv_c(R_g)}{dR_g} \right|_{R_\odot}= -1.34 \pm 0.21\ {\rm km\ s}^{-1} \ {\rm kpc}^{-1} 
\end{equation} 
measured from Cepheids \citep{mroz19}.  
%\begin{equation}  \left. \frac{dv_c(R_g)}{dR_g} \right|_{R_\odot}= -1.7 \pm 0.1\ {\rm km\ s}^{-1} \ {\rm kpc}^{-1} .
%\end{equation}
The slope  is related to the slope exponent 
\begin{equation}
\beta = \left. \frac{dv_c}{dR_g} \right|_{R_\odot} \frac{R_\odot}{V_{LSR}} = \left. \frac{dv_c}{dR_g} \right|_{R_\odot}  
\frac{1}{\Omega_\odot} .
\end{equation}
Using $R_\odot  = 8.12\ {\rm kpc}$ and $V_{LSR} = 233.3\ {\rm km/s}$, the slopes by 
\citet{mcgaugh19,li19} give $\beta = -0.059$ whereas the slope by 
\citet{mroz19} gives $\beta = -0.045$.
The slope by \citet{mcgaugh19} is measured in the range
$9 < R_g < 19$ kpc
whereas that by \citet{mroz19} is for $4 < R_g < 20$ kpc.
The slope by \citet{li19} is based on stars in the solar neighborhood.
We  adopt $\beta=-0.05$ as a compromise. 
%{\color{red}[ARP: so the McGaugh19 slope seems strictly for the outer galaxy (maybe even only $R>9$kpc?), while all radii in the results of this paper seem constrained to $7<R<9$kpc, with a bulk of the trajectories actually favouring the $R<R_\odot$ region. However, the Mroz19 data says their linear fit is valid for $4<R<20$kpc, so is more applicable.]}

The sensitivity of the gravitational potential to height above or below the Galactic plane, $z_g$, depends on the density distributions in thick and thin stellar disks,  gas disk and halo.
The recent local 3D models  \citep{barros16} are based on measurements for the
different galactic components  \citep{holmberg00,holmberg04,flynn06}.  We fit simple analytical functions to the vertical acceleration as a function of height above the Galactic plane  found from solar neighborhood K-giants that is shown in Figure 8 by \citet{holmberg04}.
We found a good fit to this curve within $z_g<750 $ pc with  a polynomial  function 
\begin{equation}
%\ddot z_g = - \frac{d\Phi_z(z_g)}{dz_g} = -\alpha_h \sqrt{|z_g|} {\rm sign}(z_g)
\ddot z_g = - \frac{d\Phi_z(z_g)}{dz_g} = -\alpha_1  z_g   -\alpha_2 z_g^2 {\rm sign}(z_g),
 \label{eqn:ddotzg}
\end{equation}
with constants
 \begin{align}
%\alpha_h &= 0.058\ ({\rm km/s})^2 \ {\rm pc}^{-3/2}
%= 1834\ ({\rm km/s})^2 \ {\rm kpc}^{-3/2}. 
\alpha_1 &= 4207.0\ {\rm Gyr}^{-2} \\
\alpha_2 & = -2792.2\ {\rm Gyr}^{-2}\ {\rm kpc}^{-1}.
\label{eqn:alpha_h}
\end{align}
This vertical acceleration is derived from a vertical potential function 
\begin{equation}
% \Phi_z(z_g) = \frac{2}{3} \alpha_h |z_g|^\frac{3}{2} + {\rm constant} .
 \Phi_z(z_g) = \frac{1}{2} \alpha_1 z_g^2 +    \frac{1}{3} \alpha_2 |z_g|^3 + {\rm constant} .
 \end{equation}
Poisson's equation applied in the mid-plane at $R_\odot$ gives a value for the frequency
 of low  amplitude vertical oscillations, 
\begin{align}
  \nu^2 &= \left. \frac{d^2 \Phi_z(z_g) }{d z_g^2} \right|_{z_g =0} =  4 \pi G \rho_0 - 2 \beta \Omega_\odot ^2
  \label{eqn:nu}
\end{align}
  where $\rho_0$ is the mid-plane mass density
  and we have used the potential of equation\;\ref{eqn:Phi_R} for the radial derivative terms.
The estimated value for the mid-plane density near the  Sun  is $\rho_0 = 0.10\ M_\odot {\rm pc}^{-3}$  \citep{holmberg04}.  This value is comparable to the value for the local density of matter found in recent work \citep{mckee2015}, which includes the density of visible stars (with improvements to prior work particularly for the density of M dwarfs and white dwarfs), the gas density, and the inferred dark matter density.   Using our adopted value for $\Omega_\odot$ and this midplane density,  the frequency of low amplitude vertical oscillations $\nu \approx  0.076$ rad/Myr  and the period of vertical oscillations is 83 Myr.
The frequency of oscillations computed using $\alpha_1$ is somewhat lower,  $\nu = \sqrt{\alpha_1}  =  0.065$ rad/Myr.
These frequencies are lower than that used by \citet{candlish14} whose Galactic models have
$\nu\approx 0.095$ rad/Myr  near the mid-plane (see their Figure 6). 
We attribute the discrepancies to the different conventions adopted for $R_\odot$ and $V_{LSR}$.

Orbits are integrated backwards using the Galactic potential model of equations \ref{eqn:Phi} -- \ref{eqn:alpha_h} and with python's general purpose integration routine \texttt{odeint} which calls
the LSODA routine from the FORTRAN77 library \texttt{odepack}. 
Each orbit consists of a series of positions and velocities as a function of time $t$ where $t=0$ is
the present and $t<0$ corresponds to  times in the past.
We integrate multiple separate orbits for each stellar association, each with slightly different initial conditions. 
The initial conditions for each orbit are the mean estimated value of the stellar association central position $x_h,y_h,z_h$
and velocity $U,V,W$  plus  randomly generated offsets in these six quantities 
that are based on estimates for the spatial extent and velocity dispersion of the association.
The initial position and velocity offsets for integration are generated using a normal distribution and standard
deviations $\sigma_x, \sigma_y,  \sigma_z,  \sigma_U, \sigma_V, \sigma_W$ 
 in the phase-space coordinates for each association. 
Because we desire estimates for both spatial extent and velocity dispersion of each association, we use
measurements for stellar associations that are
 based on  a  multivariate fitting algorithm  \citep{malo13,gagne18a}.

\section{Sample of Young Stellar Associations, clusters and moving groups}
\label{sec:sample}

Most of the young stellar associations, clusters and moving groups
 we use for this study are taken from Table 9 by \citet{gagne18a}.
This table lists  values for central coordinates and velocities $x_h,y_h,z_h,U,V,W$  
and standard deviations for these quantities
 found using the  BANYAN  algorithm \citep{gagne18a}. 
BANYAN (Bayesian Analysis for Nearby Young AssociatioNs) 
models  the distribution of stars in the young stellar associations with multivariate Gaussian 
distributions in 6 dimensional phase-space. 
The standard deviations, $\sigma_x, \sigma_y,\sigma_z,\sigma_U,\sigma_V,\sigma_W$, 
reflect the spatial extent and velocity dispersions of the associations, 
not errors in measuring these quantities. %{\color{red}[ARP:  seeing as your brought it up; I assume such errors are smaller than these dispersions and so unimportant for this work?]}

The longer orbits are integrated, the larger the errors in the orbit positions.  To mitigate this uncertainty, we
 restrict our study to associations that are younger than 70 Myr. 
We have discarded the 118 Taurus group and the Platais 8 cluster \citep{platais98} because they have been neglected
in recent studies and the constraints on their ages are poor.
To the associations listed by \citet{gagne18a} we add the Argus association, but with measurements from the BANYAN analysis using  the members and measurements by \citet{zuckerman19}  and a recently discovered  $\approx 62$ Myr old stellar association, $\mu$ Tau   \citep{gagne20_prep}.
We have checked that the results of the BANYAN analyses by \citet{gagne18a} are consistent
with the mean position and velocity measurements by other recent works \citep{binks15,riedel17,miretroig18}.

For some associations (e.g., those associated with the Scorpius-Centaurus OB association), 
more recent Gaia based observations have improved upon central positions and velocity
 dispersion (e.g., \citealt{wright18}) but have not fit the spatial extent of the association.
We have adopted not to use these more precise measurements as there are correlations
between the measured variables from the BANYAN analysis \citep{gagne18a} and we would like to integrate multiple
trial orbits for each association.   Recent studies
have uncovered additional substructure (in age, velocity and position)
 in some star formation regions, such as Corona-Australis -- \citep{galli20}, 
Taurus -- \citep{fleming20},  Scorpius-Centaurus -- \citep{pecaut16,wright18} and Orion -- \citep{kos19,tian20}.
Such substructure corresponds to gradients over small distances in the Galaxy compared to
 distances travelled since birth in the orbit.  We ignore substructure in the associations
 and star formation regions here,   
 but keep in mind that  a fuller and more accurate picture of the pattern of star formation in the Galaxy might be sensitive
to stellar association substructures.

The star formation regions, moving groups, open clusters and stellar associations used here, 
their abbreviations and their measured ages $t_{\rm age}$, are listed in
Table \ref{tab:age}.  Standard
deviations for the ages $\sigma_{\rm age}$, are estimated from the ranges and uncertainties 
given in the literature with citations
for the age estimates  also listed in this Table.  The central positions and velocities 
and the standard deviations from the BANYAN analyses are listed in Table \ref{tab:kin}. 

The youngest associations are born in at least two different filamentary extinction and molecular
cloud structures.
The Scorpius-Centaurus star formation region
includes the $\rho$ Ophiucus star formation region (ROPH), the Upper Scorpius (USCO), Lower Centaurus Crux (LCC), and 
Upper Centaurus Lupus (UCL) groups that lie above the Galactic plane and 
connect to a molecular filament that contains the Aquila Rift molecular clouds  \citep{bell15,mamajek16,pecaut16} 
at Galactic longitude $l\sim 18^\circ$ and a distance of $d\sim 200$ pc \citep{zucker20}.
%, and $\eta$ Chamaeleonti groups, and $\epsilon$ Chamaeleontis association. 
In contrast, the 
Taurus-Auriga star formation region (TAU) and 32 Orionis group (THOR) are below the Galactic plane 
and might instead be associated with the filament showing the Radcliffe wave that 
contains the Orion star formation region \citep{alves20}. 
Maps of the current locations of stellar associations are shown in Figures 4 and 5 by \citet{gagne18a}.
We will discuss association locations in context with the extinction and molecular gas filaments in more detail in subsequent sections. % when we plot their orbits in rotating frames.

\section{Results}
\label{sec:results}

Using backwards orbit integration we first look at stellar association birth sites in two-dimensions or equivalently 
projected into the Galactic plane.  In section \ref{sec:heights} we discuss birth heights above
or below the Galactic plane. In sections \ref{sec:rot} and \ref{sec:height_rot} we discuss the birth sites in 
rotating frames.  In sections \ref{sec:uvw} and \ref{sec:zvz} we discuss the stellar associations 
in context with the solar neighborhood's stellar velocity and vertical phase-space distribution.

\subsection{Estimated Birth locations}
\label{sec:birth_loc}

%{\color{red}[ARP: This section feels like it should be in Section\;\ref{sec:results}]}
Estimated birth locations and velocities computed from our orbit integrations, along with their uncertainties,
 are listed in Table \ref{tab:birth}.
Birth height, galactocentric radius and azimuthal angle,  $z_b, R_b, \theta_b - \theta_\odot$,
and birth velocity components $v_{z,b}, v_{R,b}$ 
are mean values at the association age $t_{\rm age}$ of 30 integrated orbits with randomly generated
initial conditions,  chosen as  described at the end of section \ref{sec:pot}.  We computed 
a standard deviation from the scatter of the values in the 30 orbits at $t_{\rm age}$.  
We also computed a standard deviation from a single orbit 
by weighting points in the orbit with 
 a factor that depends on the age uncertainty or spread 
\begin{equation}
w(t) = \exp\left( - \frac{(|t| - t_{\rm age} )}{2\sigma_{\rm age}^2} \right). \label{eqn:age_weight}
\end{equation}
The uncertainties for birth positions and velocities listed in Table \ref{tab:birth} are the result of summing these
two  estimated standard deviations in quadrature.  Errors caused by
the spread in initial conditions usually dominate those arising from the age uncertainty.
We neglect errors in the orbits caused by uncertainty in astronomical constants $z_\odot$, $R_\odot$, 
$V_{LSR}$, $U_\odot$, $V_\odot,$ and $W_\odot$ and parameters describing the potential model, 
$\beta,  \alpha_1$, and $\alpha_2$.

In Table \ref{tab:birth} we also list
the maximum height above or below the Galactic plane $|z|_{max}$ reached in the orbit.  We measured these from the backwards orbit integrations by integrating longer than a full vertical oscillation period.
This is a measure of the orbit's amplitude of vertical oscillations.  The uncertainty
is the standard deviation computed from the scatter  in  $|z|_{max}$ for 10 orbits with different initial conditions.   We also computed and list 
a measure of the orbital eccentricity from the maximum and minimum
radius reached in the orbit
$e = \frac{R_{g,max} - R_{g,min}}{R_{g,max} + R_{g,min}}$, with uncertainty estimated
the same way as for $|z|_{max}$.
We also list 
the radius $R_L$ of a planar circular orbit with the same $z$ component of angular momentum
computed using equation \ref{eqn:R_L}.  As the potential is axisymmetric, the $z$ component
of angular momentum per unit mass, $L$, is a conserved quantity and only depends on an orbit's initial conditions.  
The standard deviation of $R_L$  is computed
by propagating the errors in the initial conditions.    
The birth tangential velocity component $v_{\theta,b}$ can be computed
from $R_L$ and birth radius $R_b$ by inverting equation  \ref{eqn:R_L}, giving  $L = R_\odot V_{LSR} \left(\frac{R_L}{R_\odot}\right)^{\beta + 1}$ and $v_{\theta,b} = L/R_b$.

\begin{figure}
    \centering
   \includegraphics[width=3.6in, trim={0mm 0mm 0mm 0mm},clip]{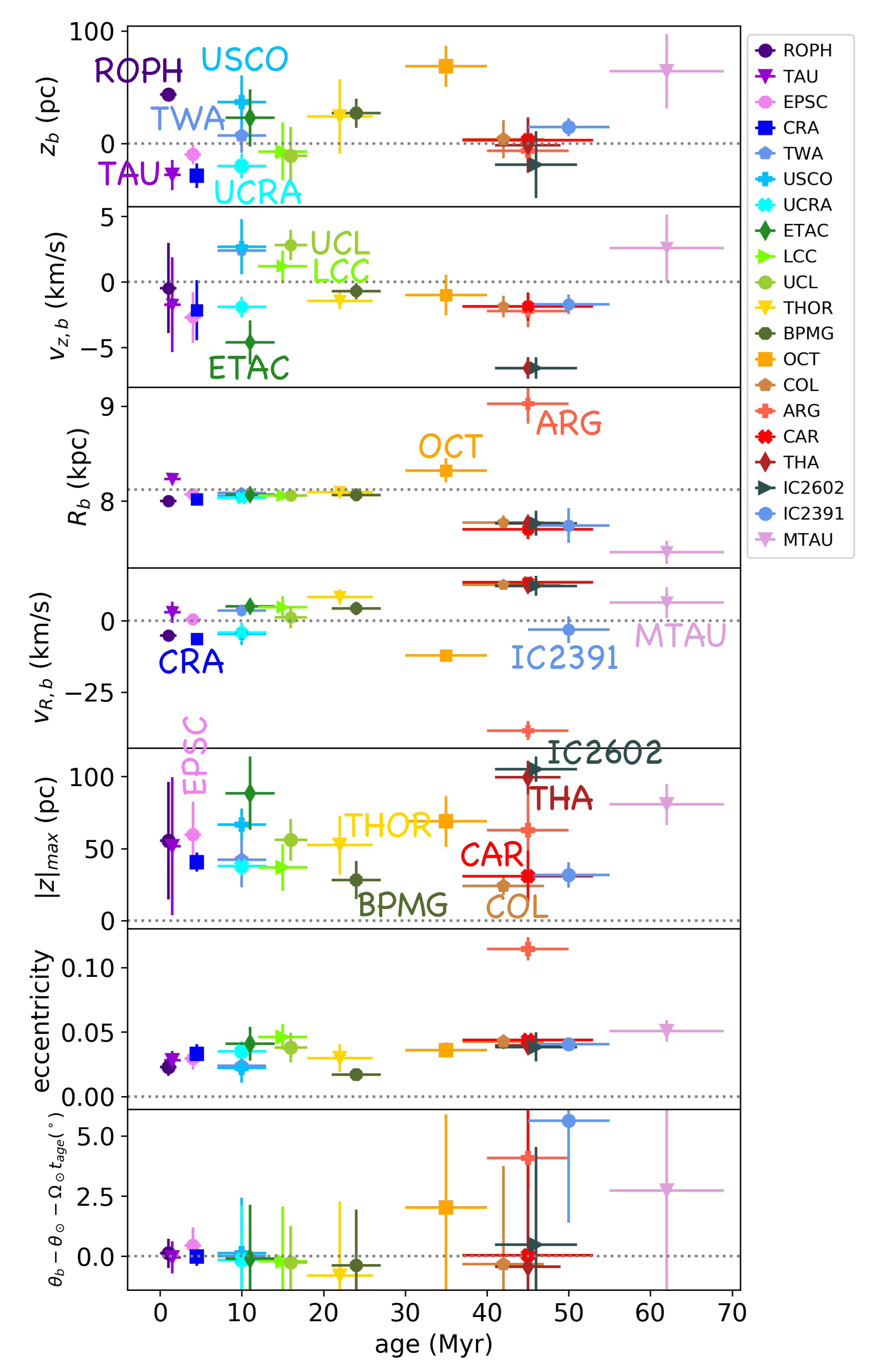} 
\caption{Birth locations and velocities of stellar associations plotted as a function of their age.
Top to bottom panels show birth height $z_b$ (in pc), vertical component of velocity at birth
$v_{z,b}$ (in km/s), birth galactocentric radius $R_b$ (in kpc), radial component of velocity at birth $v_{R,g}$
(in km/s), 
maximum height reached in the orbit $|z|_{max}$ (in pc), eccentricity, and angle in a frame rotating with
the local standard of rest (in degrees).  The $y$ coordinates of points plotted here
are listed in Table \ref{tab:birth}.  The $x$ coordinates are listed
in Table \ref{tab:age}. Vertical errors bars are uncertainties due to spread
in the initial coordinates and velocities.   Horizontal error bars  are uncertainties due to 
age spread or uncertainty. %{\color{red}[ARP: $v_{\theta,b}$ absent? I also wonder if there is anything interesting to see if different marker types indicated their types (e.g. associations, groups)]}
    \label{fig:birth1} } 
\end{figure}

Birth locations and velocities and maximum orbital height are plotted as a function
of stellar association age in Figure \ref{fig:birth1}.  In this plot the 
vertical error bars are  uncertainties due to the spread
in the initial (and current) positions  and velocities.   Horizontal error bars show  
age spread or uncertainty.  
Horizontal coordinates  are listed in Table \ref{tab:age} and vertical coordinates are listed in Table \ref{tab:birth}.
In Figure \ref{fig:birth1} the horizontal grey lines are at a vertical coordinate of zero except in the fourth panel where it is
at the galactocentric radius of the Sun, $R_g = R_\odot$.
The bottom panel in Figure \ref{fig:birth1} shows  the angle $\theta_b - \theta_\odot  - |\Omega_\odot t_{age}| $
in degrees.  This angle gives birth azimuthal angle in a frame corotating with the local standard of rest.
This angle is the difference between the birth angle 
$\theta_b$ and that of a particle in a circular orbit that is  at the location of the Sun at $t=0$.

Figure  \ref{fig:birth1} shows some trends with age. The youngest stellar associations 
are born both above and below the Galactic plane and are on nearly circular orbits.
Intermediate age associations (20 - 30 Myr) have lower vertical amplitudes (lower $|z|_{\rm max}$).
The older associations are coming into the solar neighborhood from both larger and smaller radii.
These trends will be discussed in more detail below.

\begin{figure*}
    \centering
   \includegraphics[width=5.0in, trim={0mm 0mm 0mm 0mm},clip]{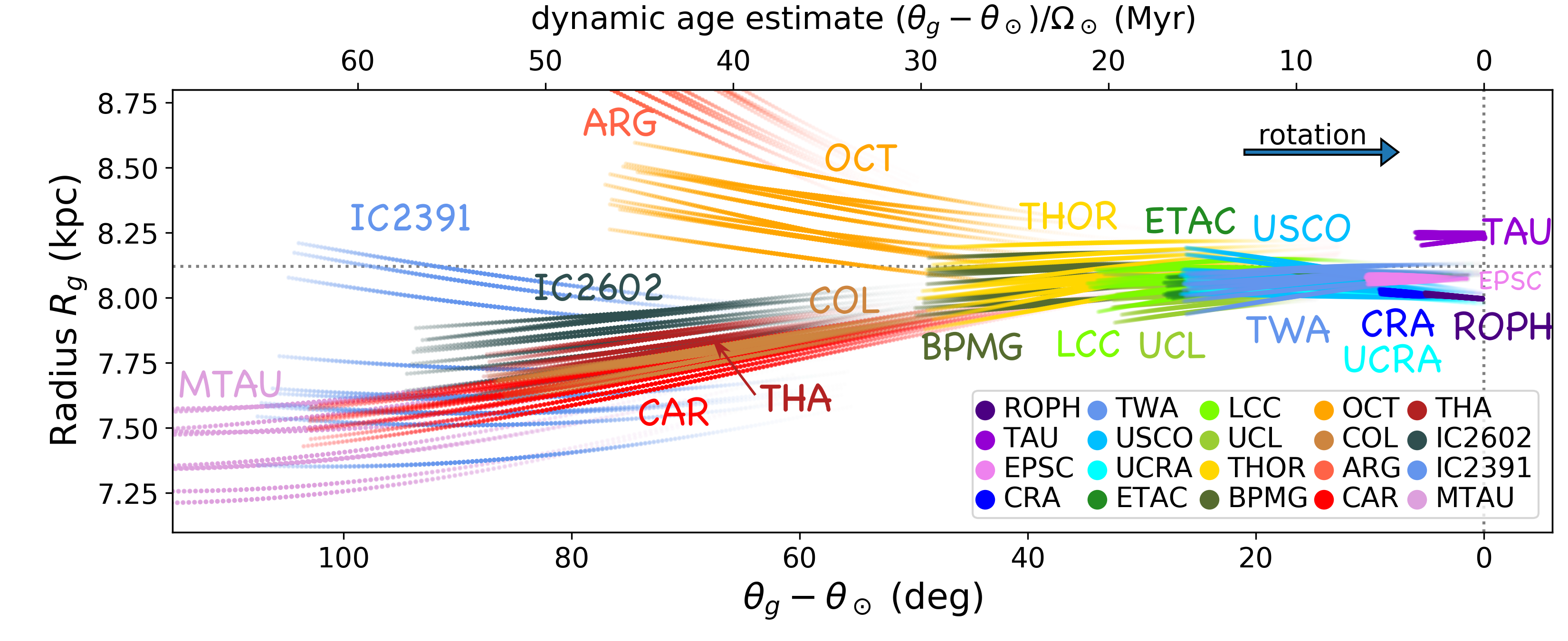}\\ % annotated
\caption{Backwards integration of young stellar associations younger than 70\ Myr that are listed in 
Table \ref{tab:age}. 
We plot orbits as a function of galactocentric radius and azimuthal angle from that of the Sun, $\theta - \theta_\odot$
 in degrees.  
Ten orbits are shown for each stellar association to illustrate how the orbits depend
upon the spread in initial conditions.  Each stellar association is shown with a different color point and 
the colors are labelled  in the key.
The opacity of the points is only high for points at times near the estimated stellar association birth age.
Points on the left side of the plot are older associations that were born further away from the Sun.
The ages of these points can be estimated using the angular rotation rate of an object
in a circular orbit at the galactocentric radius of the Sun, $(\theta_g - \theta_\odot)/\Omega_\odot$.
This age coordinate is shown with the top axis.
Associations were plotted in order of seniority, oldest ones first.
Most of these associations were born at lower galactocentric radius
and moved radially outward into the solar neighborhood where they are found today.
%Outliers to the trend shown by the other associations are the Octans and Argus stellar associations  and that associated with IC 2391.  
Dotted grey lines intersect at the location of the Sun and the direction of Galactic rotation
is shown with an arrow.  Note that the azimuthal angle increases to the left.
    \label{fig:rtheta_all}}
\end{figure*}

We have checked that variations in the adopted value of  
galactocentric solar radius $R_\odot$, local standard of rest velocity   $V_{LSR}$ 
and rotation curve slope $\beta$,  within the errors of recent measurements, do not significantly 
affect the morphology of the orbits in Figures \ref{fig:birth1} and \ref{fig:rtheta_all} or trends discussed below.  
With a flat rotation curve corresponding to exponent $\beta=0$,  the birth radii of the oldest associations  
 at somewhat smaller (a few hundred pc lower) galactocentric radii.

\begin{figure*}
    \centering
\ifshort
  \includegraphics[width=7.0in, trim={0mm 0mm 0mm 0mm},clip]{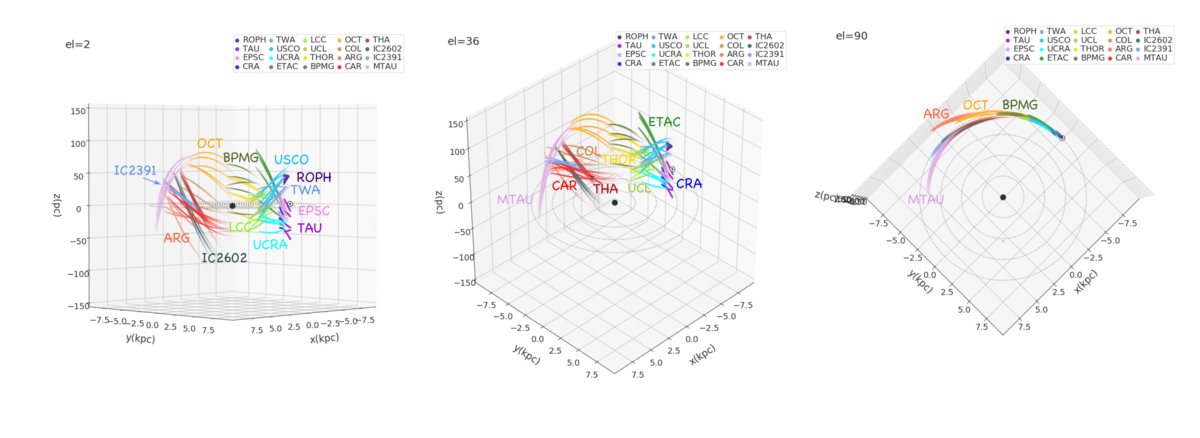}
\else
   \includegraphics[width=7.0in, trim={0mm 0mm 0mm 0mm},clip]{movieb.png}\\ % annotated!
 \fi
\caption{3D plots of backwards orbit integrations.  This figure is similar to Figure \ref{fig:rtheta_all}
but five orbits are shown per stellar association.   The Galaxy center is the black dot.
The position of the Sun is a shown with a Sun symbol. The leftmost plot shows the heights of the
backwards integrations whereas the rightmost plot shows the orbits projected onto the galaxy midplane. 
The central plot show an intermediate view.  We include a link to a supplemental movie that is available online {\bf movie3D.gif}
of the 3D plot with a varying viewing angle.
\label{fig:movie3D}}
\end{figure*}

\subsection{Birth locations in the Galaxy} % 4.2
\label{sec:birth}

In Figure \ref{fig:rtheta_all}, 
we show the galactocentric radius and azimuthal angle $\theta_g - \theta_\odot$ for each of 10 orbits for each stellar association.  The orbits of each association are shown in a different color 
 with colors identified in the key.  
At time $t$,  the opacity of the point is weighted by the weight function $w(t)$ (equation \ref{eqn:age_weight}) 
which peaks at 1 when time $|t|$ is the age of the association and has width dependent
on the association age spread. %{\color{red}[ARP: I'm a little puzzled by this, why do most lines have smoothly decaying opacity to the right but seem to have a harder edge to the left in Fig2? Was some hard cut-off (e.g. $3\sigma_{\rm age}$) used for backwards integration time?]}. 
This way points are only visible
in the plot near the estimated association birth location.
%The standard deviation $\sigma_a$ is the uncertainty in the age or in some cases the estimated association age spread.  
%The values we used  for $t_a, \sigma_a$ are listed in Table \ref{tab:age} along with citations describing their measurements.
Points on the left side of the plot are older associations that were born further away from the Sun.
The ages of these points can be estimated using the angular rotation rate of an object
in a circular orbit at the galactocentric radius of the Sun, $(\theta_g - \theta_\odot)/\Omega_\odot $.
This approximate age in Myr is shown with the top axis in Figure \ref{fig:rtheta_all}.

The $x$ axis in Figure  \ref{fig:rtheta_all}, showing Galactic azimuthal angle,
 is reversed so that Galactic rotation is to the right.  We chose this convention
so that the plots can more easily be compared to maps of Milky Way spiral structure, extinction and molecular clouds.
The direction of Galactic rotation (clockwise) is shown with an arrow on the top right.
The assumed galactocentric radius of the Sun is marked with a horizontal dotted grey line. 
The azimuthal angle of the Sun is marked with a dotted vertical line, so the current position of the Sun is on the right-hand side of the plot where the two dotted grey lines cross.
In Figure \ref{fig:rtheta_all}, increasing galactocentric $R_g$ upward along the $y$-axis increases the heliocentric $y_h$ coordinate.   Moving to the right along the $x$ axis in Figure \ref{fig:rtheta_all}  increases  the heliocentric coordinate $x_h$.  

Views of the integrated orbits plotted in 3D are shown in Figure \ref{fig:movie3D}.   We provide a supplemental movie {\bf movie3D.gif} of the 3D plot seen at varying viewing elevations that is available online.

\subsubsection{Inwards or outwards radial motion after birth} % 4.2.1
%\label{sec:4.2.1}

We find that most of the stellar associations have moved outwards radially from their birth locations. 
In other words, their current galactocentric radius exceeds their birth galactocentric radius. 
Exceptions to this trend are very young associations such as Corona Australius (CRA) and $\rho$ Ophiucus 
(ROPH) associations  that are still near 
 their birth clouds.   Among the older associations, 
the Octans  (OCT) and Argus (ARG) associations have orbits that differ
from the other associations.  These two are also exceptions because 
 they have moved inward to reach the solar neighborhood since their birth.

As the rotation curve is nearly flat, the epicyclic frequency $\kappa \sim \sqrt{2} \Omega$ 
with $\Omega$, the angular rotation rate of a particle in a planar circular orbit.
This gives an epicyclic oscillation period
of about 155 Myr at radius $R_\odot$.  For the Argus and Octans associations, their ages correspond to only about a quarter
of an epicyclic oscillation period.  This implies that they must have been moving radially inward soon after birth, rather
than outwards after birth as are most of the other associations.

The distance moved radially, or equivalently the epicyclic amplitude or orbital eccentricity, is largest for
the Lower Centaurus Crux group (LCC), and Argus (ARG) and $\mu$-Taurus (MTAU) associations.  
If we assume that these associations were born in spiral arms, then the parent spiral features 
 caused a greater degree of non-circular motion and so were probably 
more massive than the parent features of the youngest associations. 

\subsubsection{Expectations for inward or outward motion from models and simulations} % 4.2.2
\label{sec:4.2.2}

Spiral features in 
N-body simulations usually show pattern speeds similar to or lower than the local
angular rotation rate $\Omega$ (e.g., \citealt{quillen11,grand12,kawata14}). %\footnote{Exceptions include  spiral features that are driven by a bar (e.g., \citealt{lindblad96}) and spiral features that strengthen due to interference between patterns (e.g., \citealt{quillen11,comparetta12}).} 
%might cause a feature or pattern to travel faster than $\Omega$. 
%
Spiral arms in N-body simulations usually exhibit lower spiral pattern speeds at larger radii (e.g., \citealt{quillen11,grand12,kawata14}).
As the Octans and Argus associations were born at larger Galactic radius, perhaps their birth
arm had a slower pattern speed.   The interstellar medium, with a sound speed similar to
10 km/s, is shocked as it passes over a spiral arm
(e.g., \citealt{shetty07,dobbs10,pettitt15,shu16}).
The shock compresses the gas, increases the gas density and lowers the gas velocity
in the frame moving with the spiral pattern.  The compressed gas 
should have an  angular rotation rate that is approximately 
the same as that of the spiral pattern.    In other words, in the frame moving with the spiral
arm, the molecular clouds should have low angular rotation rate (e.g., \citealt{dobbs10}).
Consider a star born with a low radial velocity component. 
If the star has an angular rotation rate that is slower than that of a particle in a circular orbit at the same radius then its angular momentum is lower than that of the particle and it would 
 move radially inward after birth. 
Perhaps the Octans and Argus
associations were born in spiral features with pattern speeds that are lower than the angular rotation  rate of a circular orbit,
$|\Omega_s| < |\Omega|$, at their birth radius.  The opposite could be true of most of the rest of the associations if
$|\Omega_s| > |\Omega|$ for their birth spiral arm.

\citet{dobbs10} compared simulations of different galactic morphologies to assess their impact on the spread of cluster ages, inferred from the locations of densest gas elements (rather than a specific sub-grid star formation prescription). These models included a flocculent galaxy, %a barred galaxy, 
a galaxy with a steady spiral and a tidally perturbed galaxy. 
A steady spiral pattern
shows a gradient in the ages of recently formed stars across each spiral arm. 
However if star formation
not only occurs along the arm, but in spurs and features emanating from the arm, the gradients are
shallower. Spiral arms caused by tidal perturbations are more complex, exhibiting positive or negative age gradients across spiral arms. 
%Galaxy-companion interactions can involve multiple out of plane encounters that make the dynamics considerably more complex than a single prograde, in-plane, fly-by (e.g., \citealt{pettitt17}).
The simulated flocculent galaxy by \citet{dobbs10} shows localized bursts of star formation.  
Because there is no simple trend in age vs birth locations and kinematics (see also Figure \ref{fig:birth1}), 
continuous star formation without spurs or armlets along a single steady spiral arm seems ruled out.
The other scenarios could be consistent with the stellar association birth locations and kinematics.
Unfortunately we have not found a study measuring young star epicyclic phases as a function of age from simulations (whether moving radially inward or outward) but perhaps this additional information could in future help differentiate
between spiral structure models. 
%{\color{red}[ARP: we could easily try this I think! See email.]}.
 
\begin{figure*}
    \centering
   \includegraphics[width=3.0in, trim={0mm 0mm 0mm 0mm},clip]{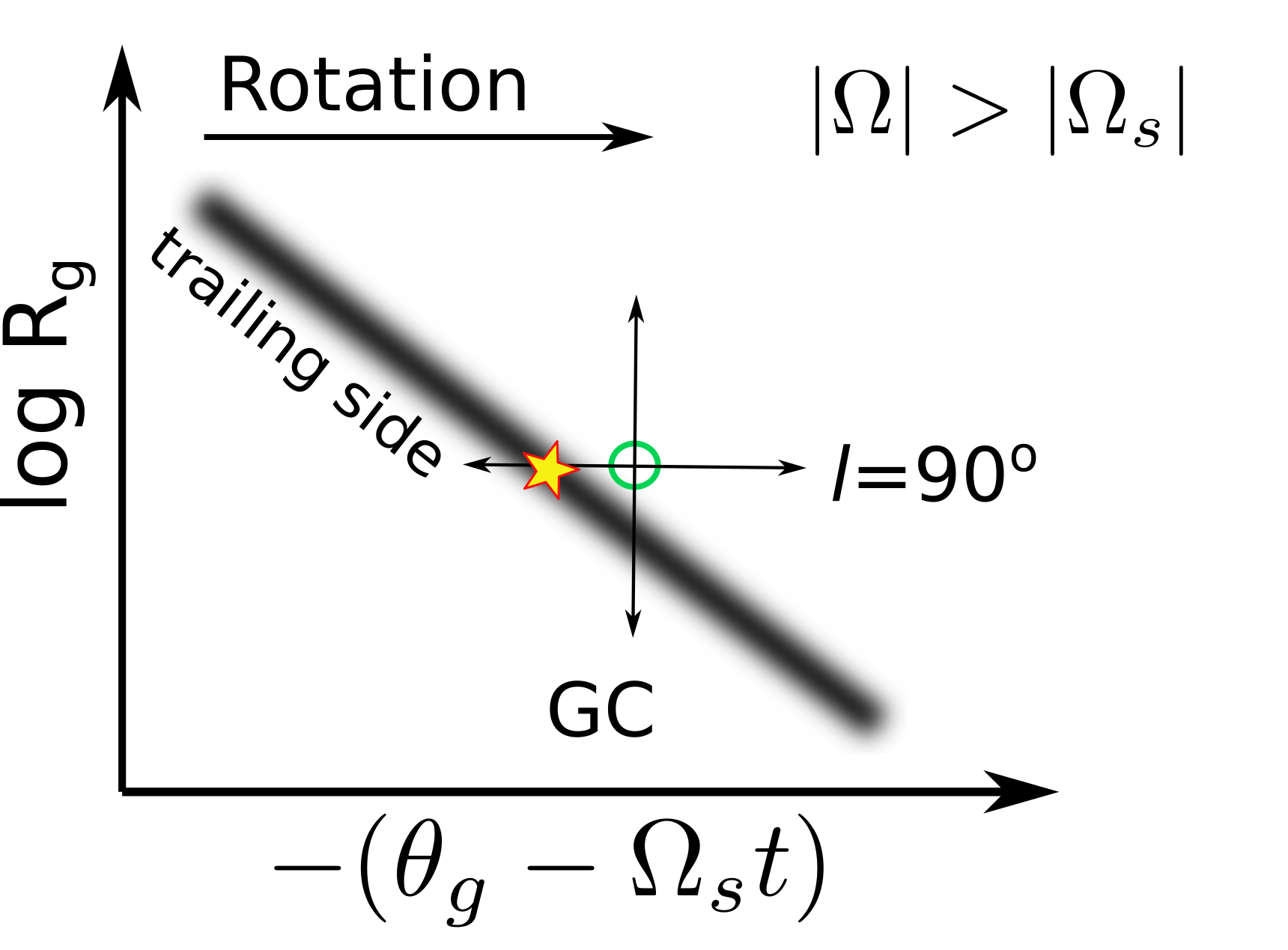}  %cartoons
   \includegraphics[width=3.0in, trim={0mm 0mm 0mm 0mm},clip]{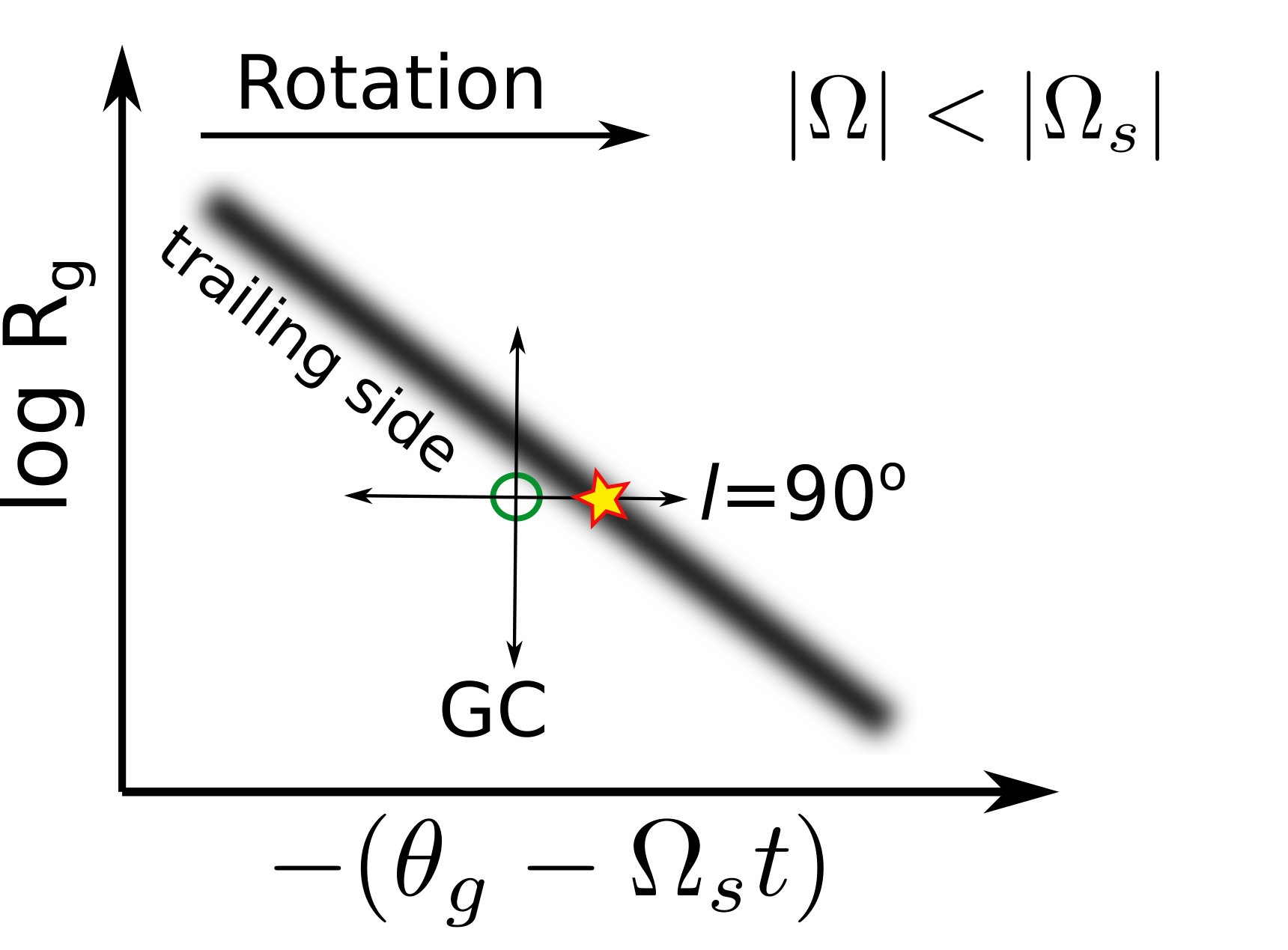}\\
\caption{We illustrate a logarithmic spiral arm. 
Here the $y$ axis is log galactocentric radius and the $x$ axis is galactic azimuthal angle in 
the frame moving with pattern speed of $\Omega_s$.
Clockwise galactic rotation is to the right and shown with the arrows on top.  These arrows do
 not necessarily show the direction of motion in the rotating frame.  
We show a trailing logarithmic spiral arm (equation 
\ref{eqn:logspiral}) with a wide diffuse black bar.  The arm's pitch angle determines the slope of the bar
with negative slope corresponding to a trailing arm.
a) If the angular rotation of a circular orbit is faster than the spiral pattern speed,  $|\Omega |> |\Omega_s|$,  stars born on the arm peak (shown as a yellow star) pass the spiral arm leaving 
them to the right of the spiral arm (shown as an open circle).   If the Sun is located at the open circle 
the 4 arrows show different viewing directions with  Galactic longitude of $90^\circ$ in the direction of rotation
and GC representing toward the Galactic  center. 
%In the rotating frame of the spiral arms, stars born on the arm peak would then move to the right.
b) If the angular rotation of a circular orbit is slower than the spiral pattern speed,  $|\Omega |< |\Omega_s|$, stars born on the arm peak would move to the left of the arm.
    \label{fig:trailing}}
\end{figure*}

\subsubsection{Location of stars after birth on leading or trailing sides of a parent spiral arm}

To illustrate on which side of an arm stars would be located, we show in Figure \ref{fig:trailing}  a trailing logarithmic spiral arm in frames rotating
with the pattern speed of the arm.  In Galaxies, the angular rotation rate is often higher at smaller radius than in the outskirts.  A linear radial feature that winds up due to differential rotation would exhibit a `trailing' spiral.   Leading spiral features are probably rare \citep{buta03}.
%(they could be driven by a recently formed bar or caused by a strong tidal perturbation). %%% citations?
Figure \ref{fig:trailing}a illustrates the case with arm pattern speed moving
slower than the local rotation, $|\Omega_s |< |\Omega|$ and 
Figure \ref{fig:trailing}b illustrates the opposite case.
In polar coordinates a logarithmic spiral arm pattern can be described with peak at galactocentric radius 
$R_{\rm peak} (\theta_g,t)$ 
where 
\begin{equation}
 \alpha_s \ln \left( \frac{R_{\rm peak}(\theta_g,t)}{R_{s0}} \right)  = \theta_g - \theta_\odot - \Omega_s t . \label{eqn:logspiral}
 \end{equation}
At time $t=0$, the current time, 
the arm has a peak at galactocentric radius $R_{s0}$ and at angle $\theta_\odot$.
The arm pitch angle is $p  = {\rm arctan}\ \alpha_s^{-1}$ and its pattern speed is $\Omega_s$.
As we have adopted a coordinate system giving clockwise Galactic rotation,  
 $\dot \theta <0$,  a trailing arm has winding angle $\alpha_s>0$. The pattern  
moves in the same sense as  rotation, so the pattern speed  $\Omega_s <0$.
A logarithmic trailing arm is linear on a plot of $\log R_g$ vs
$\theta_g$.  The arm's pitch angle determines slope of the arm on this illustration 
with negative slope corresponding to a trailing arm.

In the illustration of Figure \ref{fig:trailing},  galactic rotation is to the right,
however in the rotating frame, stars move to the right in Figure \ref{fig:trailing}a 
and to the left in Figure \ref{fig:trailing}b. 
With stars exceeding the pattern speed, as shown on in Figure \ref{fig:trailing}a, stars born on the arm pass  the arm and are located to the right of the arm, and near the direction of Galactic longitude $l \sim 90^\circ$. 
With pattern speed exceeding that of that of a circular orbit,
the pattern moves faster than the stars.  In a frame rotating with the pattern, the arm is fixed and the stars move in the opposite direction and to the left, 
as shown in Figure \ref{fig:trailing}b.

%We discuss these radial motions in the context of peculiar velocities, with respect to galactic rotation, that is seen in simulations that contain spiral structure. 

N-body simulations that exhibit spiral structures that are approximately corotating with the local angular rotation rate, $|\Omega_s| \sim |\Omega|$, have  
tangential peculiar velocities that are slower on the trailing side and faster on the leading side 
of a spiral arm and radial peculiar velocities that point outward on the trailing side and inward on the leading side \citep{grand15,baba16}. 
When its tangential velocity is slower than that of a circular orbit at the same radius, a star has lower angular momentum than the circular orbit  
and so must spend most of its orbit at lower radius.
If the spiral pattern moves slower than the local angular rotation rate $\Omega$ then the associations
are currently found on the leading side of the arm, as shown in Figure \ref{fig:trailing}b.    
Stellar associations with birth radius lower than their current values are consistent
with the outward radial velocities and sub-circular birth tangential velocity reported in the simulations
by \citet{grand15}, on the trailing side of approximately corotating arms, if the spiral pattern speed
is slightly higher than $\Omega$.  If the spiral arm or arms in which Octans and Argus associations
formed has the opposite relation, $|\Omega_s| <| \Omega_\odot|$, then the associations are currently 
on the leading side of the arm and the trends noted by \citet{grand15} would be consistent
with birth site exterior to $R_\odot$, as we have observed from their orbits. 

%In contrast,  test particle simulations  tend to show inwards radial motion near the arm peak and outwards radial motion in the inter-arm regions \citep{siebert12}. 

\subsubsection{Birth on leading or trailing sides of a spiral arm}
 
Stars and gas in proximity to a spiral arm can gain or lose angular momentum due to the torque exerted
by the gravitational pull of the spiral arm.  
When the spiral arm is approximately corotating with the galactic rotation, 
the change in angular momentum of nearby stars and
gas clouds is larger because they remain on one side of the arm longer  
(e.g., \citealt{kawata14}).   Stars and gas clouds  trailing the arm, and stars born in these clouds, 
gain angular momentum and would then move 
outward in radius, whereas those leading the arm lose angular angular momentum and  would 
move inward  \citep{kawata14}.    An alternative explanation for the few associations that have moved radially 
inward after birth is that they were born on the leading side of a spiral arm that decreased their 
angular momentum rather than in a spiral arm that has a slower pattern speed than corotation. 
In this scenario,  we might expect that stars 
are born on both leading and trailing sides of arms.   If the Octans and Argus associations were
born on the leading side of an arm, then we could look for stars that were born at the same time
and in the same arm but on the trailing side.   These would be moving outward from their birth site
so would not be near the Sun, but might they be near enough to find in a deeper survey of young stars. 
We estimate the birth galactocentric radius of the Octans association at 8.3\,kpc so trailing arm birth counterparts 
to the the Octans association might
be less than a kpc away from the Sun. 

The youngest associations include Corona-Australis association (CRA) and $\rho$-Ophiucus star formation  (ROPH) 
 that are currently moving radially inward and 
were born moving radially inward.  The Taurus-Auriga star forming region (TAU) is currently moving outward
and was born moving radially outward.   However, it is unlikely that these were born on opposite sides
of the same arm because they are still in proximity to their birth clouds and they are at different heights.

In summary, birth sites for most of the stellar associations are interior to the Sun's galactocentric radius
and moved outward after birth.
This would be consistent with birth in a spiral arm with pattern speed that is higher than $\Omega_\odot$
placing these  associations currently on the leading side of their birth arm.
This expectation follows from birth in a shock 
moving with the spiral arm and so with higher  angular momentum than a circular orbit. 
The direction of motion is consistent with 
 peculiar velocities seen in simulations of approximately
corotating transient spiral structures by \citet{grand12}. 
Alternatively  the associations that moved outward were born on the trailing side of a corotating spiral
arm that increased their angular momentum through its gravitational torque  \citep{kawata14}.
The Octans and Argus associations are exceptions as they were born outside $R_\odot$
and this suggests that their parent spiral arm has a pattern speed slower than $\Omega_\odot$.
Alternatively they could have been born on the leading side of a nearly corotating spiral feature
and pulled inward by the arm itself \citep{kawata14}.  
Other scenarios, such as involving tidally excited spiral structure or spurs and armlets 
extending from strong arms might also
account for these inferences (e.g., \citealt{dobbs10}).

\subsection{Birth Heights}
\label{sec:heights}

\begin{figure*}
    \centering
   $$ \begin{array}{cc}
   \includegraphics[height=4.3in, trim={5mm 10mm 5mm 0mm},clip]{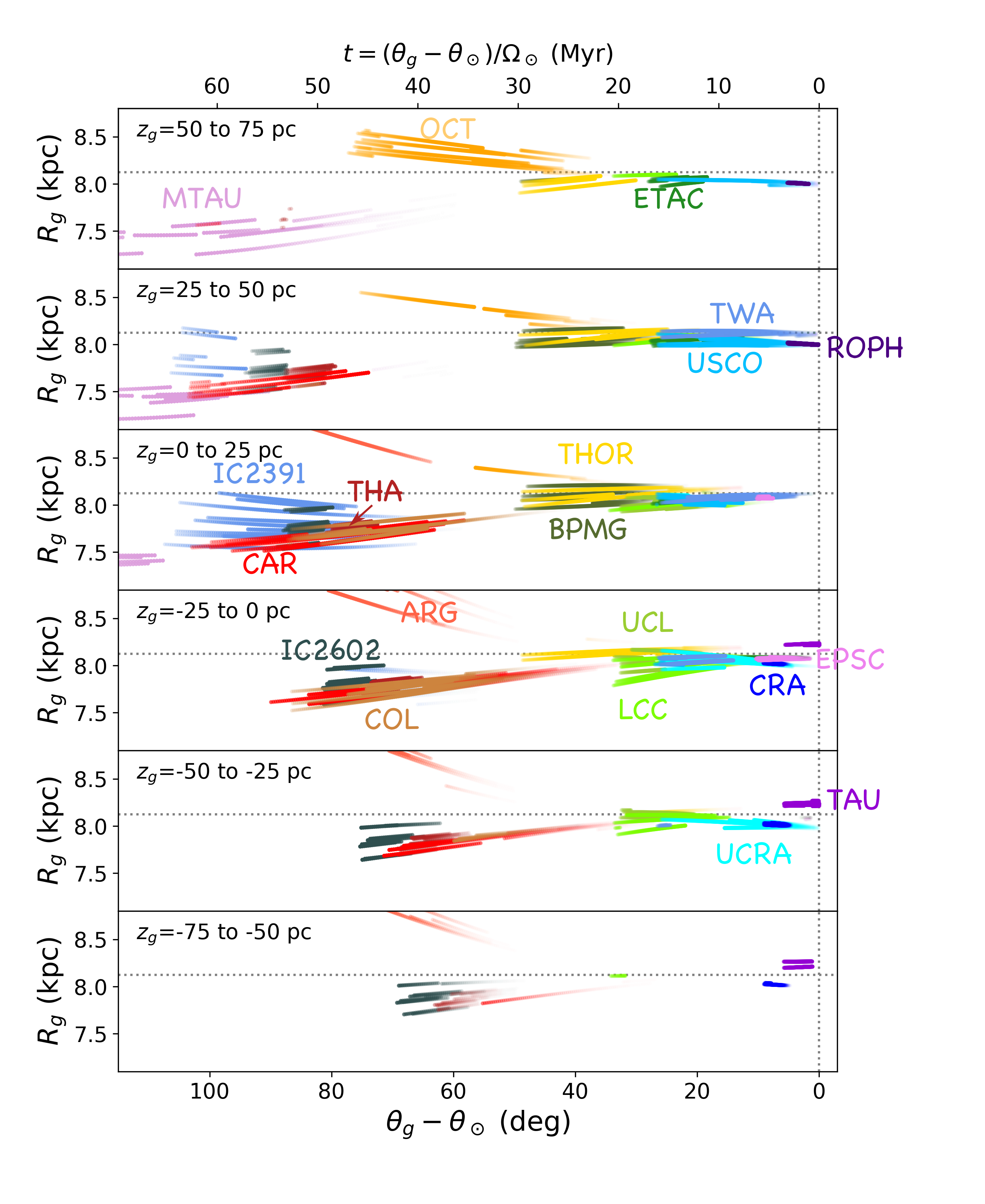} &
    \includegraphics[height=4.3in, trim={5mm 10mm 5mm 0mm},clip]{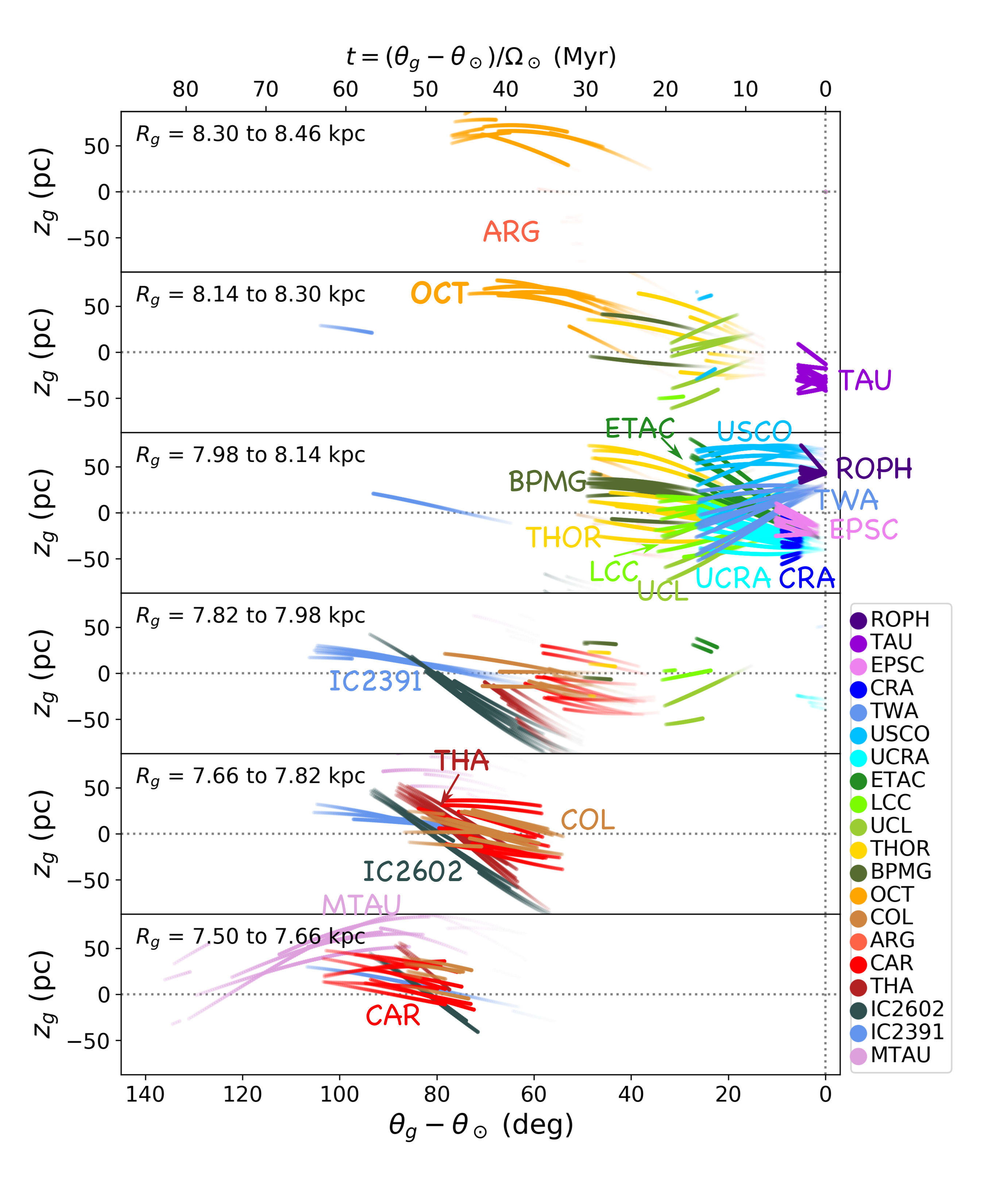} 
     \end{array}$$
\caption{a) Similar to Figure \ref{fig:rtheta_all}  except we  
plot galactocentric radius $R_g$ vs angle for stellar association orbit positions that lie at different heights above or below the Galactic plane.
Each panel corresponds to a different range of heights, and these are labelled on the top left of each panel.
An orbit that varies in height would have points in more than one panel.
b) We plot  height $z_g$ vs angle and
each panel only plots points in different ranges of galactocentric radii. 
The radial ranges are plotted on the top left of each panel.
The older clusters are on the left
and the younger ones on the right.
These figures suggest that 
recent star formation has taken place in a corrugated molecular disk.
    \label{fig:z4}}
\end{figure*}

We  examine the estimated stellar association birth heights listed in Table \ref{tab:birth}. 
Figure \ref{fig:z4}a is similar to Figure \ref{fig:rtheta_all} and shows young stellar association orbits 
except we only plot orbit points 
in different $z_g$ ranges.    Each panel  corresponds to a planar slab with width 25 pc.  The vertical slab 
upper and lower $z_g$ values 
are labelled on each panel.  Figure \ref{fig:z4}b is similar to Figure \ref{fig:z4}a except
the $y$ axis in each panel is $z_g$ and each panel only shows points in different ranges
of galactocentric radius. %{\color{red}[ARP: it would be purely cosmetic, but I think a 3D plot in simple $(x,y,z)$ space would look pretty cool to show their orbtal paths!]}

%Among the youngest star formation regions that formed below the Galactic plane, the Taurus-Auriga (TAU) star formation region is at a larger radius than the Corona-Australis (CRA) one.  
The intermediate-age associations ($\sim 20$ Myr) such as the $\beta$-Pictoris moving group (BPMG) and 
the 32 Orionis group (THOR) were born near the Galactic plane and at low eccentricity (also see Figure \ref{fig:birth1}).  
The Octans (OCT) and Argus (ARG)
associations were probably born above the Galactic plane at larger radii than $R_\odot$. 
The Columba  (COL), Tucana-Horologium  (THA) and the Carina (CAR) associations 
were probably born near the Galactic plane.  
%associations and IC 2602 were probably born above the Galactic plane, whereas IC 2391 and the Carina association (CAR)
%were probably born near or below the Galactic plane.  
These inferences are based on a fixed and axisymmetric Galactic potential, so could be updated or
corrected for orbits integrated in more complex potential models.

The birth locations of similar age associations that are at different heights are also at different radii and angles.  The youngest associations can be divided into three groups, 
those at lower radius and above the Galactic plane (ROPH, USCO, TWA), 
those at lower radius and below the Galactic plane (CRA, UCL, LCC, UCRA)
and those at larger radius and below the Galactic plane (TAU, EPSC).
Figure \ref{fig:z4} does not necessarily imply that more than one molecular filament must exist
simultaneously at the same $R_g, \theta_g$ but at different heights 
above or below the Galactic plane.  Transitions in the birth sites of the different age
associations seen in Figure \ref{fig:z4} suggest that there are spatial variations in
the parent molecular cloud distribution.
Recent star formation could have taken place in a corrugated molecular disk, as suggested
by the current distribution of molecular clouds  near the Sun \citep{zucker20,alves20}. 

We have checked that the patterns shown in Figure \ref{fig:z4} are present 
with 20\% higher or lower values of parameter $\alpha_1$ which we have used to 
describe the vertical acceleration in the Galaxy  (see equation \ref{eqn:ddotzg}).
Setting the height of the Sun $z_{g,\odot}$ to 20 pc does not significantly alter the overall appearance of the orbits.

Stars born above or below the Galactic plane, or with non-zero vertical velocity components, will undergo vertical oscillations.   The maximum heights 
reached above or below
the Galactic plane measured from the backwards orbit integration
 are listed in Table \ref{tab:birth},  and plotted as a function of stellar association
age in the fourth panel of Figure \ref{fig:birth1}. 
The youngest stellar associations (less
than 10 Myr old) have a range of maximum heights $|z|_{max}  \sim $ 20 to 60 pc (see Figure \ref{fig:z4}b). 
However the intermediate age  association $\beta$ Pictoris moving group (BPMG) 
and 32 Orionis (THOR) moving groups, with age $\sim 20$ Myr, 
the Columba and Carina associations (COL and CAR; age $\sim 45 $ Myr) have lower vertical
amplitudes, less than 30 pc.
The associations with the largest vertical amplitudes tend to be the older ones.
The Octans and Argus associations ($\sim 40$ Myr old) that were born at larger radius, have maximum heights
in the range of 35 to 60 pc.  The Tucana Horologium and $\mu$ Taurus associations  (THA, MTAU) and  IC2602 
have the highest maximum heights in the range 60 to 100 pc.

The wavelike or undulating structure seen in the filament of molecular clouds associated
with the Local arm \citep{alves20} could be comprised of clouds undergoing similar
amplitude vertical motions but at different phases of oscillation.  Alternatively there might be spatial variations in
the amplitudes of the vertical motions.  The dip in the vertical amplitudes (the $|z_{\rm maz}|$-age plot in Figure \ref{fig:birth1}) of the intermediate age
  32-Orionis and  $\beta$ Pictoris moving groups suggests that there are spatial variations in the vertical amplitudes of the parent molecular clouds.
Tightly wound bending waves that travel through the disk  (e.g., \citealt{hunter69}) would 
be expected to have amplitudes that are slowly varying with galactocentric radius and angle.
In contrast, a phase wrapping model, where the Galactic disk was perturbed in the past and 
passively evolved afterward, could give a disk exhibiting variations in vertical amplitude over short distances \citep{delavega15}.
A tidal perturbation on the disk would excite stars in one region of the Galaxy
more than other regions. Stars or gas clouds from different locations while the perturbation occured could
be near the Sun now (e.g. \citealt{candlish14,delavega15,darling19}).  An additional possibility is that these structures arise from orbits of stars moving in a dark matter halo that significantly departs from spherical symmetry due to past mergers with dwarf galaxies.  However, it is likely that such an effect is minimal close to the mid-plane as prior work studying the evolution of halo shapes finds that baryonic-dominated regions are nearly spherical \citep{debattista08,prada2019}.

% The very young associations were born above and below the Galactic plane, but with some (like $\rho$ Oph; ROPH) born above the galactic plane and others  born below it (like the Corona-Australis star-forming region' CRA).
%Associations like $\beta$ Pictoris (BPMG) that are about 20 Myr old were born near the Galactic plane. At about 30 Myr the Octans association (OCT) was  born above the plane at larger radii than $R_\odot$  whereas the Columba and Carina associations were born within  the solar circle and near the Galactic plane. The Tucana Horologium (THA) association and IC 2602 were born slightly within $R_\odot$ and above the plane 40-50 Myr ago.  This figure also supports a picture where recent  star formation has taken place in a corrugated molecular disk.

In summary, trends in the birth heights of the stellar associations suggest that
they were born in a corrugated disk of molecular clouds.   
Maximum vertical heights or vertical amplitudes reached above or below the galactic plane 
can be high (50--100 pc) for the youngest and oldest stellar associations
but seem to be somewhat lower, only $\sim 30$ pc for those 20--30 Myr  old.    
This implies that there were spatial variations in 
the amplitudes of the vertical motions in the parent molecular cloud distributions.

\begin{figure*}
    \centering
      \includegraphics[width=5.0in, trim={0mm 0mm 0mm 0mm},clip]{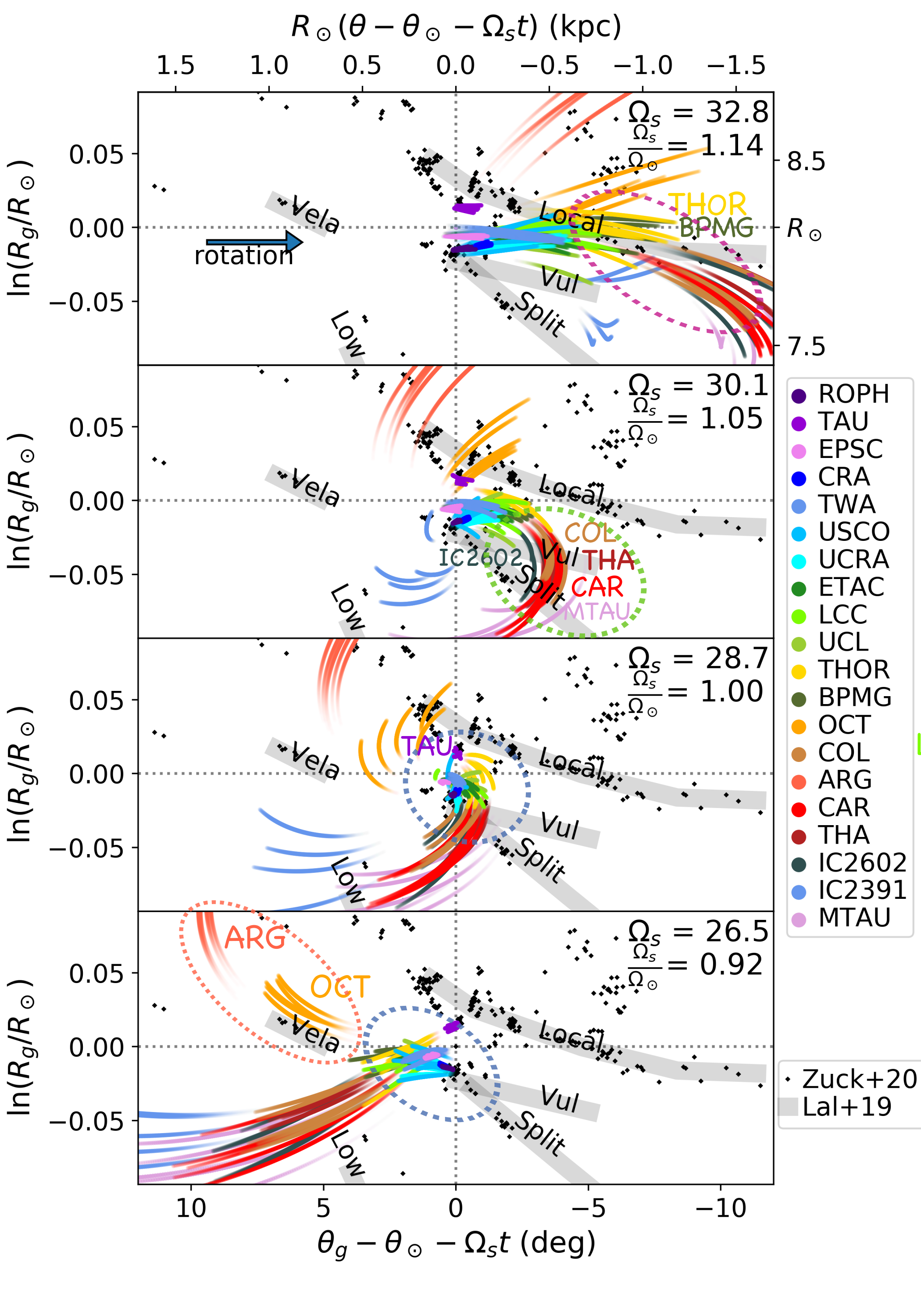}\\
\caption{Backwards orbit integrations of young stellar associations plotted in rotating frames.
Each panel is similar to Figure \ref{fig:rtheta_all} except 
we plot galactocentric radius versus azimuthal angle in a frame rotating with a spiral arm pattern. 
The assumed pattern speeds for each panel are labelled on the top right of each panel
in units of km~s$^{-1}$~kpc$^{-1}$ and in units of $\Omega_\odot$.
To give a sense of scale for the $x$ axis,
the top axis shows distances in kpc along the solar circle of radius $R_\odot$. 
The $y$ axis is the natural log of galactocentric radius and we have marked
three values of radius in kpc on the right hand side of the top panel to provide a sense of scale.
The third panel from the top has a corotating pattern speed, $\Omega_s = \Omega_\odot$.
A spiral pattern near corotation allows the younger stellar  
associations to have been born in a compact region (marked with a dotted blue oval).  
If the pattern speed is near but below corotation (bottom panel, dotted blue oval), the young stellar
associations  could have been born on a continuation of the extinction feature known as the `Split'.
The Argus (ARG) and Octans (OCT) associations could
have been born in a trailing filament with a lower pattern speed (the bottom panel, dotted orange oval). 
The Columba (COL), Tucana-Horologium (THA), Carina (CAR), $\mu$-Taurus (MTAU)
associations and IC2602 cluster could have been
born in a trailing filament with a faster pattern speed (see top two panels and the dotted green oval).
If the pattern speed is as fast as 1.14 $\Omega_\odot$, the same feature could
have been the birth site of the $\beta$ Pictoris (BPMG) and 32 Orionis (THOR) moving groups
(see the dotted pink oval in the top panel).
No single pattern speed can account for all stellar associations. 
    \label{fig:lom_all_new}}
\end{figure*}

\subsection{Birth locations in rotating frames and birth spiral arm candidates} % 4.4
\label{sec:rot}

Keeping in mind that different associations were born at different heights, we now discuss orbits
in different rotating frames.  We explore the possibility
that molecular cloud filaments near the Sun are spiral features, moving
as waves,  in which stellar associations were born.

In Figure \ref{fig:lom_all_new} we show backwards orbit integrations for the same
sample of stellar associations as in Figure \ref{fig:rtheta_all}, however, instead of plotting orbit positions
at prior times $t$ 
as a function of galactocentric angle, we plot association positions as a function 
of angle $\theta_g - \theta_\odot - \Omega_s t$ where $\Omega_s$ is an assumed pattern speed.
Each panel shows the stellar associations in a frame that is rotating with 
 a different possible pattern speed.
On the top axis we show the distance along the solar circle, $R_\odot (\theta_g - \theta_\odot - \Omega_s t)$  to give a sense of scale for the $x$ axis.  
The $y$ axis is the natural log of galactocentric radius instead of radius but we have marked
three values of radius in kpc on the right hand size of the top panel to provide a sense of scale.
A logarithmic spiral arm would be linear on Figure \ref{fig:lom_all_new}.
Trailing spiral arms should have a negative slope on these plots (as shown 
in Figure \ref{fig:trailing}).
%

%we compare the distribution of stellar association birth locations in various rotating frames  to molecular cloud positions and extinction features. 
In Figure \ref{fig:lom_all_new}
extinction features, based on those labelled in 3D local extinction maps by  
\citet{lallement19}, are plotted as grey bars.  Molecular clouds 
are plotted as black dots using the database by \citet{zucker20}.
Unfortunately many of the masers identified in the feature
called the `Local Spur' at Galactic longitude 
$l \sim 50^\circ$ by \citep{xu16,xu18}  are further than 2 kpc away from the Sun and  outside the region spanned by Figure \ref{fig:lom_all_new}.  
The Local Spur may be connected to an extinction feature at $l \sim 50^\circ$ and distance
from the Sun $d<500$ pc labelled as 
'Vul' by \citet{lallement19} (see their Figure 14).  
A filament denoted the `Split' by \citet{lallement19}
contains the Serpens molecular clouds at longitude $l \sim 18$ to $30^\circ$ and distance $d \sim 500$ to 1200 pc,
%above the Galactic plane  $z_g \sim 50  $ pc,
the Aquila Rift at $l\sim 18^\circ$ and a distance of $d\sim 200$ pc, 
and connects to the nearby Scorpius-Centaurus star formation region.  
The Vela C cloud at $l \sim 256^\circ$ and $d \sim 900$ pc  is also prominent in the extinction  maps.
The extinction  filament associated with the Local Arm contains 
the Orion star formation region (at $l \sim 200^\circ$ and $d\sim 400$ pc),
Cepheus Near (at $l \sim 110^\circ$ and  $d \sim 340$  pc), 
North America (at $l \sim 84^\circ$ and  $d \sim 800$  pc) and 
Cygnus X  clouds (at $l \sim 80^\circ$ and  $d \sim 1000$ pc).
The longitudes and distances given here 
for these molecular clouds are based on those listed in Table A1 by \citet{zucker20}. 
There is a nearby or lower component to the Sagittarius Carina arm denoted 
`Low' in figure 14 (for Lower Sagittarius Carina arm) by \citet{lallement19} 
that is approximately at $l \sim 330^\circ$ and $d \sim 1$ kpc.
These are the extinction features that are shown as grey bars and labelled on Figure \ref{fig:lom_all_new}.

%{\bf  moved from section 4.2}
As discussed in section \ref{sec:4.2.2},  we suspect that associations 
that have moved radially outward after birth 
would have been born on spiral arms with pattern speed $|\Omega| < |\Omega_s|$ 
and they should be currently on the trailing side of their birth spiral arm. This implies 
that that a counterpart to their birth arm would currently be located on the sky in the direction of Galactic 
rotation or near Galactic longitude $l \sim 90^\circ$ or on the right in Figure \ref{fig:lom_all_new}. 
Candidates for the birth arm of the Octans and Argus associations, born with $R_g > R_\odot$, 
 would be in the opposite direction,  such as that associated with the Vela C cloud at  $l = 265^\circ$. 
 
Can the Octans (OCT) and Argus (ARG) associations have been born on the same arm?
If that arm has a faster pattern speed than $\Omega_\odot$, then a quite open rather than tightly wound
arm would be required to parent both of these associations.
The Octans and Argus associations only lie on the same tightly wound arm if that
arm has a slower pattern speed, $|\Omega_s| \la |\Omega_\odot|$ (see bottom panel of Figure \ref{fig:lom_all_new}, and marked with a dotted orange oval) and agreeing with our discussion in section \ref{sec:birth}. 
If the Vela C cloud is their parent filament  then
its pattern speed $|\Omega_s| \sim 26.5 \ {\rm km\ s}^{-1}\ {\rm kpc}$ which is lower than 
$\Omega_\odot$, as would be expected from the Octans and Argus association birth radii.
We note that the birth filament could have been corrugated as the birth heights of the Octans and Argus
associations differ ($z_b \sim 70, -6$ pc, respectively).
 
\begin{figure*}
    \centering
%    $$\begin{array}{ccc}
  \includegraphics[height=4.0in, trim={0mm 0mm 0mm 0mm},clip]{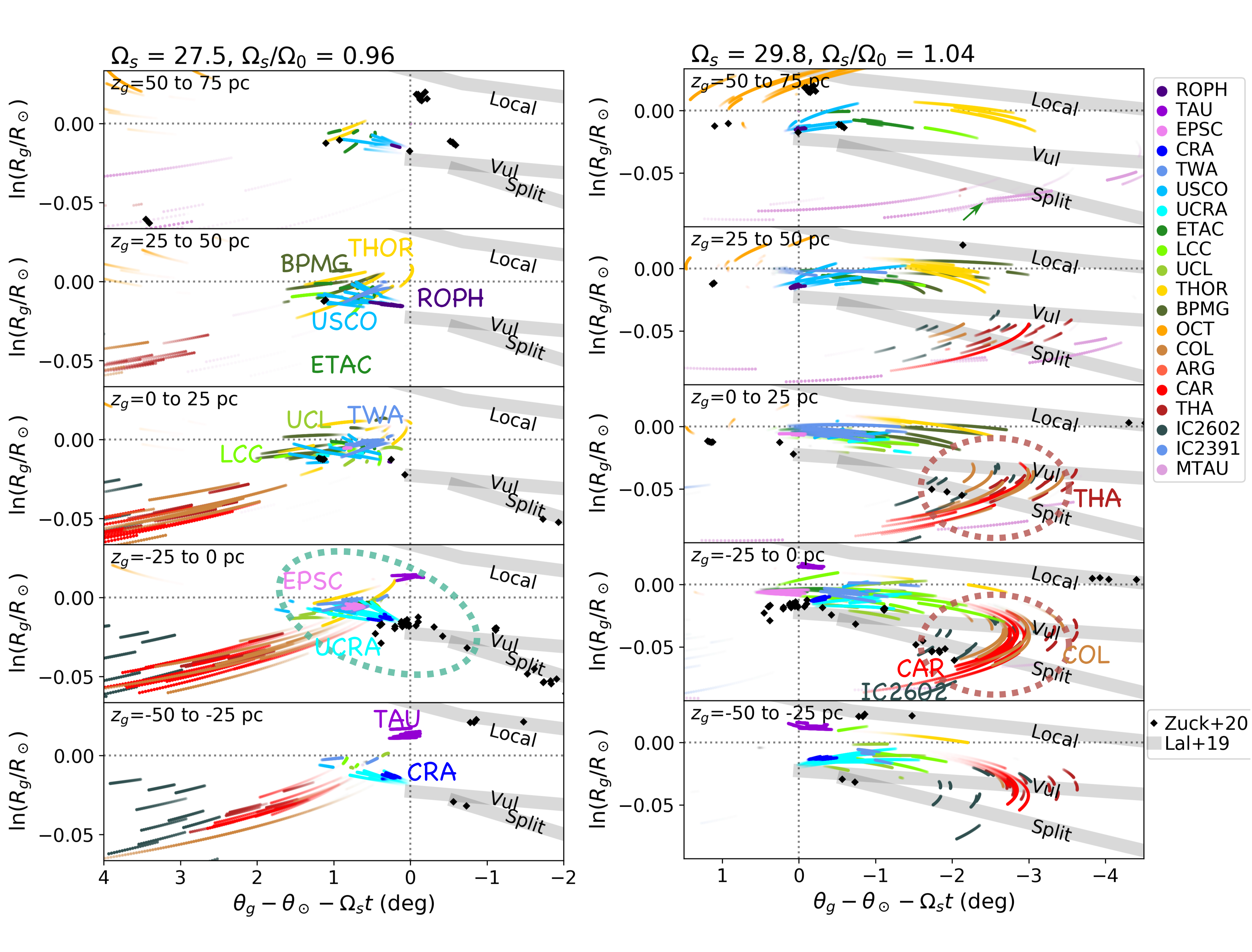} 
%     \end{array}$$
\caption{Orbits in rotating frames. Similar to Figure \ref{fig:lom_all_new} except each set of panels shows shows a single pattern speed and in each panel we 
only plot positions that lie in a single planar slab. 
The range in height $z_g$ is denoted on the top left of each panel.
The assumed pattern speed $\Omega_s$  is written on the top of each set of panels
in units of km s$^{-1}$ kpc$^{-1}$ and $\Omega_\odot$.  
The younger associations span a range of heights and are unlikely to have been
born on a single corrugated filament. 
The birth arm of the younger associations could have been thick, or contained sub-filaments at different
heights. 
The Columba (COL), Tucana-Horologium (THA), and Carina (CAR) associations 
have similar birth heights and locations in the rotating frame (see brown dotted ovals on the right set of
panels), so they could
have been born on the same filament that is associated with the `Split' or `Vul'
extinction features.   
    \label{fig:lomz4}}
\end{figure*}

For the young associations (younger than 20 Myr old), 
candidates for the birth arm could be an extension of the local arm at 
Galactic longitude $l \sim 90^\circ$ or the local spur at $l \sim 50^\circ$ (connected to the extinction feature denoted `Vul' here; \citealt{xu18}), or the extinction features denoted  the `Split' at  $l \sim 30^\circ$ \citep{lallement19}.  If the pattern
speed of the birth arm is near corotation  $\Omega_s \approx \Omega_\odot = 28.7 \ {\rm km\ s}^{-1}\ {\rm kpc}^{-1}$, then the associations were born in a compact region in the rotating frame 
(see second panel from the bottom in Figure \ref{fig:lom_all_new}
and dotted blue oval) but in between the Local arm and the Split. 
The near corotation pattern speed would support some recent estimates of local Galactic spiral pattern speeds (e.g, \citealt{naoz07,dias19}).
If the pattern speed was $|\Omega_s| \approx  26.5 \ {\rm km\ s}^{-1}\ {\rm kpc}^{-1}$
then the young associations were born on a filament that is an extension of the `Split' extinction feature.
(Note this slower pattern speed would contradict our expectation
 that associations that have moved outward after birth were born in faster pattern speed filaments.)
The Taurus-Auriga star formation region could be associated with
the Local arm rather than an extension of the Split, however it is not at the same height as 
the clouds that appear nearest to it on Figure \ref{fig:lom_all_new} (this can be seen  
 in Figure \ref{fig:lomz4}).  

With a faster pattern speed, $|\Omega_s| \sim 30\ {\rm km\ s}^{-1} {\rm kpc}$, 
as shown on the second from top panel of Figure \ref{fig:lom_all_new} (and marked with dotted green oval), many of the intermediate age associations
(the Columba (COL), Tucana-Horologium (THA), Carina (CAR), $\mu$-Taurus (MTAU)
associations and IC2602 cluster but not the IC 2391 cluster) could have been born in a single spiral feature.
The faster pattern speed would be consistent with their lower birth radii.   
With a fast enough pattern (top panel and dotted pink oval),  the youngest associations (less than 20 Myr old) are near a line with
positive slope on Figure \ref{fig:lom_all_new} (in the top panel), corresponding to a leading rather than trailing pattern.   This means that 
a filament with a fast pattern speed $|\Omega_s| \ga 30 \ {\rm km\ s}^{-1}\ {\rm kpc}$ is unlikely to have
been the birth site of both the intermediate age associations and the youngest ones. 

Spiral features in 
N-body simulations usually show pattern speeds similar to or lower than the local
angular rotation rate $\Omega$ (e.g., \citealt{quillen11,grand12,kawata14}).   Exceptions include 
spiral features that are driven by a bar (e.g., \citealt{lindblad96}) and spiral features that tilt and strengthen due to interference between patterns (e.g., \citealt{quillen11,comparetta12}).   A pattern with a fast pattern
of $\Omega_s \approx \ga 33 \ {\rm km\ s}^{-1} {\rm kpc}$ would allow associations such
as the $\beta$ Pictoris (BPMG) and 32-Orionis (THOR) moving groups, the Columba (COL), Tucana-Horologium (THA), Carina (CAR), $\mu$-Taurus (MTAU)
associations, and the IC2602 cluster to be born on the same filament 
(Figure \ref{fig:lom_all_new}  top panel).
However the birth radius of the 45 Myr old Carina (CAR) and Tucana-Horologium (THA) 
associations is $R_b \sim 7.7 $ kpc where
the angular rotation rate $|\Omega| \sim 30.3 \ {\rm km\ s}^{-1}\ {\rm kpc}$. 
The faster pattern speed of $33 \ {\rm km\ s}^{-1} {\rm kpc}$ exceeds the angular rotation rate of their birth radii.
If their birth filament had such a fast pattern speed then the spiral arm dynamics eould
be in one of these exotic categories.

%In Figure \ref{fig:lom_all_new}, the higher pattern speeds put the birth locations of the  younger associations between the local arm filament and the Split and Vul filaments seen in dust extinction maps.
%However a spiral pattern near corotation (the third panel from top) allows the younger  associations to have been born in the same arm that is an extension of Split or Vul filaments.
%However, a single molecular filament would not  be consistent with the scatter in birth heights in the youngest associations.
%Two nearby nearly corotating molecular filaments, one associated with Split or Vul filaments and the other near the Taurus-Auriga association, at different heights, could have been birth sites for most of the younger stellar associations.  
%At slower pattern speeds, (the bottom panel), the Argus and Octans associations could have been born in a filament that is related to the Vela C molecular cloud complex.
%None of the pattern speeds clearly link current molecular cloud filaments to the  birth sites of the older associations such as the Carina (CAR) and 32 Orionis (THOR) associations.

Figure \ref{fig:lom_all_new} implies that in the past 50 Myr the pattern of spiral structure cannot be decomposed into a set of a few steady moving filaments.  Transient behavior and multiple molecular filaments seem necessary to explain even 
the most recent history of star formation near the Sun.

\subsection{Heights in rotating frames}
\label{sec:height_rot}

Using equation \ref{eqn:nu}, the vertical oscillation period near the Sun is about 83 Myr.  A quarter
period is only 20 Myr. This implies that the stellar associations that are younger than 20 Myr
right now have heights that are fairly near their birth heights.   Likewise, 
their birth molecular cloud heights would not have changed significantly since their birth.
This means we can compare the past birth heights of the younger associations
in rotating frames to the current positions of molecular clouds.

Figure \ref{fig:lomz4} is similar to Figure  \ref{fig:lom_all_new} except 
 in each panel we 
only plot positions that lie within specific planar slabs.  
The height ranges (in $z_g$) of each slab are printed on the top left of each panel.
%Figure \ref{fig:lomr4} is  similar to Figure \ref{fig:lomz4} except  the $y$ axes are height and in each panel we  only plot positions that lie within a range of radius.  
Each set of panels  shows a single pattern speed, one just above $\Omega_\odot$
and the other just below $\Omega_\odot$.
%The molecular cloud positions by \citet{zucker20} are plotted as black dots in each panel in  Figures \ref{fig:lomz4} and \ref{fig:lomr4}. 

% TAU can't be on the local arm because it's height is off from the nearest ones
In Figure \ref{fig:lomz4} left panels we show a pattern
speed slightly below corotation.  Because they are young, the positions of 
the youngest associations on this plot are not strongly sensitive to the pattern speed.  
Some groups related to the Scorpius Centaurus star formation region (USCO and ROPH)
are above the Galactic plane and were born above the Galactic plane
(see top panel in Figure \ref{fig:birth1}).  There are some
molecular clouds currently near their birth locations,  however there
are more molecular clouds below the Galactic plane (see Figure \ref{fig:lomz4}, left set of panels, 
green dotted oval).
%The corotating patern speed middle set of panels in Figure \ref{fig:lomr4} shows
%molecular clouds at two heights in the $R_g = 7.98$  to 8.14 panel.
We had hoped to estimate how the height of a spiral feature varied in the last
few million years, but these plots do not clearly pick out specific current molecular cloud counterparts
for the different young stellar associations or a clear pattern of up and down motions in their star
formation history.   
Most of the young stellar associations
could have been formed on an arm that is a continuation of the `Split' extinction feature, particularly
if the pattern speed is slightly lower than corotation (see Figure \ref{fig:lomz4}, left set of panels
and as discussed in the previous section).
%However the stellar association birth sites have a range of heights.  
We don't clearly
see clear evidence for birth on different filaments at different radii and heights.
The birth arm of the younger associations could have been thick, or contained sub-filaments at different
heights or been so tightly corrugated that its history is not clearly visible in our plots.

On the right set of panels in Figure \ref{fig:lomz4} we show heights for a pattern
speed slightly above corotation.  
The Columba (COL), Tucana-Horologium (THA), and Carina (CAR) associations 
have similar birth heights and locations in the rotating frame, so they could
have been born on the same filament.  
%This filament could be an extension of the `Split' extinction feature (Figure \ref{fig:lomz4} dotted brown ovals).  
These associations
are about 40 Myr old, so their vertical height would have undergone  
half an oscillation period since birth.  Their current (rather than birth) heights
are near the Galactic plane.   
At their birth locations in the rotating frame, 
some of the clouds that are associated
with the 'Split' extinction feature are also near the Galactic plane (see the second
from bottom panel in Figure \ref{fig:lomz4}, dotted brown ovals).   This implies that these associations
could have been born in an extension of the `Split' extinction feature
and their birth heights and the current heights
of molecular clouds in this filament are approximately consistent with this scenario.  

%These associations are currently further below the Galactic plane than their birth heights.
%as the $\beta$ Pictoris (BPMG) and 32-Orionis (THOR) moving groups , 
%$\mu$-Taurus (MTAU)
%In summary, many stellar associations could have been born on one or two  nearly corotating filaments that are associated with nearby dust extinction features.  
%A candidate for the current location of this arm
%is the extinction featured labelled the 'Split' by \citet{lallement19}.
%The Octans and Argus associations could have been born on the same arm. If its pattern speed were slower than $\Omega_\odot$. Its  current counterpart could be the Vela C cloud.    Multiple arms with different pattern  speeds and heights seem required to account for the stellar association birth locations.
%This suggests that   that spiral arms or molecular cloud filaments exhibited transient behavior in the past 50 Myr near the Sun.

\subsection{Stellar associations in comparison to the local velocity distribution} 
\label{sec:uvw}

 \begin{figure}
    \centering
\ifshort
     \includegraphics[width=3.4in, trim={0mm 0mm 0mm 0mm},clip]{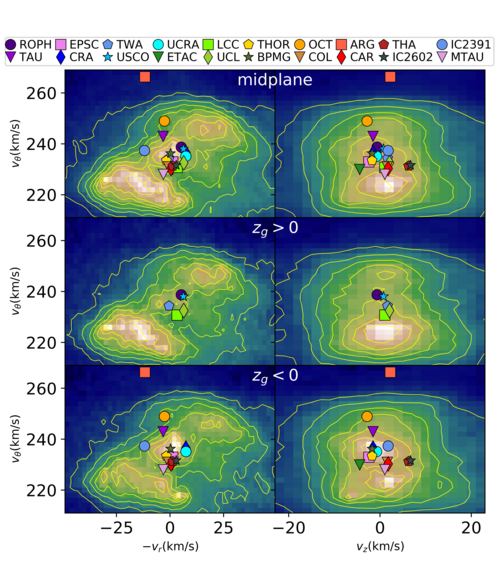}
\else
   \includegraphics[width=3.4in, trim={0mm 3mm 2mm 4mm},clip]{uvw_gaia.png} 
\fi    
\caption{Stellar association velocity components plotted as points on the velocity distribution for
nearby stars within 200 pc of the Sun using the Gaia DR2 sample.
The vertical axes are the tangential velocity component $v_\theta$ in km/s. 
For the  panels on the left,
the horizontal axes are -1 times the radial velocity component $-v_R$ (adopting sign convention for the $UV$ plane).
For the panels on the right, the horizontal axes are the vertical velocity component $v_z$.
The stellar associations tend to be located between the density peaks in the velocity distribution. 
The middle row is similar to the top row but only stars and associations above the Galactic plane are plotted.
The bottom row is similar to the top row but only stars and associations above the Galactic plane are plotted.
    \label{fig:uvw}}
\end{figure}

We discuss the stellar associations in context with the velocity
distribution %.  and vertical phase-space distributions of stars  that are present 
in the solar neighborhood.  In this section we use the current, not birth, stellar association coordinates. 

\citet{quillen18} proposed that under-dense arcs in local velocity distributions separate stars that have recently crossed and been more strongly perturbed by a particular arm from those that have not. % {\color{red}[ARP: maybe a sentence or two for the reader briefly summarising this hypothesis?]}. 
%A boundary or locus in a local velocity distributions could separate stars that have recently crossed and been more strongly perturbed by a particular arm from those that haven't.
Since stellar associations
could have  been born in a nearby arm, we can test this hypothesis with them.  They might be more likely to lie near the locus or  underpopulated region in the velocity distribution.  Stars on one side of the locus would not cross an arm, stars on the other side would cross it, and stars that graze the arm would lie on the locus.  Stars born on the arm might be near or on  the locus.

In Figure \ref{fig:uvw} we show stellar association velocity components plotted as points on top of the velocity
distribution (shown with a colour map) that is generated from nearby (within 200 pc) Gaia DR2 stars with radial velocity measurements \citep{katz18}.   The bottom two rows in Figure  \ref{fig:uvw} are similar to  the top row except
only stars and associations above or below the plane are used to make the plot.
This figure was
generated with the same database, selection criteria and numerical scripts as the figures previously presented by \citet{quillen18}. 
The axes for the left panels are galactocentric tangential and radial velocity components $v_\theta, -v_r$
and those on the right panels are $v_\theta$ and $v_z$.  Each stellar association is plotted with a different
point shape but their colours and plotting order are the same as we have used in our previous figures.

Stellar associations are expected to have been born in spiral features.  This implies that their current
velocities  should be on loci separating orbits that cross an arm from those that do not cross the same arm.
Figure \ref{fig:uvw} shows that the stellar associations studied here tend to be located between
 peaks in the local velocity distribution.  The peak at $(v_\theta,v_r) \sim (220,0) \ {\rm km~s}^{-1}$
is often called the Pleiades moving group or stream and that at $(v_\theta,v_r) \sim (240,0) \ {\rm km~s}^{-1}$
often called the Coma Berenices moving group or stream  (following the names used by \citealt{dehnen98}).
These streams contain stars with a wide range of ages (e.g., \citealt{dehnen98}).
The Coma Berenices moving group is more prominent in stars 
below the Galactic plane \citep{quillen18_galah,monari18}.
The interpeak locations of the stellar associations
on the $v_\theta, v_r$ velocity distribution is 
consistent with the hypothesis that dips in the velocity distribution are 
associated with orbits that touch nearby spiral density features.

Are the spiral patterns proposed by \citet{quillen18} consistent with the stellar association
locations on this plot? Only 6 patterns were shown (and listed in their
Table 1) and only two of them overlap
with the galactocentric radius range covered by our Figure \ref{fig:lom_all_new}.  
Because multiple features are probably required to account for the stellar
associations birth sites (as discussed in section \ref{sec:rot}), the model by \citet{quillen18} is too simplistic to be predictive.
%However, we can test the proposal that under-densities  in the velocity distribution are associated with star formation regions in spiral arms. 
The locus separating the Coma Berenices and Pleaides moving
groups was connected to pericenter with an approximately corotating arm that could be the Local Spur.
Many of the young stellar associations on Figure \ref{fig:uvw} are near this under-density 
and have moved outward in radius after birth, 
so could be consistent with this interpretation, though as discussed in sections \ref{sec:rot}
and \ref{sec:height_rot}, a filament with a single pattern is unlikely to account for all of them.

\citet{quillen18} connected 
the locus separating the Coma Berenices moving group from the Sirius/Usra Major moving
groups (at $(v_\theta,v_r) \sim (250,-20) \ {\rm km~s}^{-1}$)   
 to apocenter with the Local arm with pattern speed 
slower than corotation $\Omega_s = 27$ km s$^{-1}$ kpc$^{-1}$.
On Figure \ref{fig:uvw}, 
the Taurus-Auriga (TAU) star formation region is located on the opposite
side of the Coma Berenices moving group peak compared to the other young associations.
This would be consistent with our hypothesis that it was not born on the same
filament as the others.   The Octans association also seems located near this under-density,
however we suspect that it was born in an arm with an even lower pattern speed.
The large difference between the $v_\theta$ of Argus and Octans associations suggests
that they were not born on the same arm.

The associations that currently are above the Galactic plane (Figure \ref{fig:uvw} middle panels) are 
young groups related to the Scorpius-Centaurus star formation region. 
The rest of the associations are currently below the Galactic plane. 
Other than this, we do not see any obvious trends in the comparison between stellar association 
and velocity distributions above and below the Galactic plane in Figure \ref{fig:uvw}. 

\subsection{Stellar associations in comparison to the stellar vertical phase-space distribution}
\label{sec:zvz}

\begin{figure}
    \centering
\ifshort
    \includegraphics[width=3.4in, trim={0mm 0mm 0mm 0mm},clip]{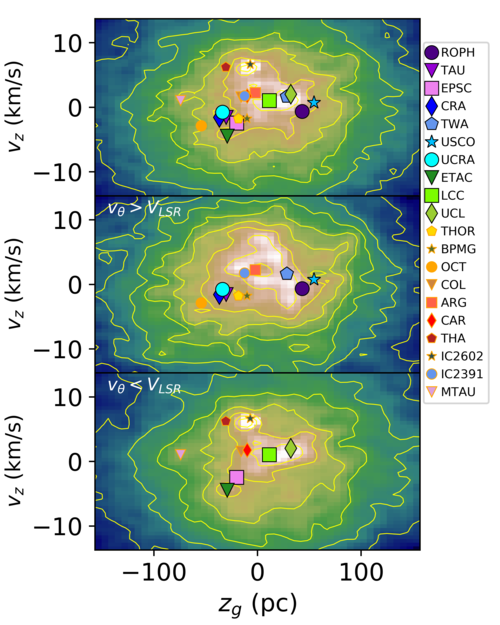} 
\else
   \includegraphics[width=3.4in, trim={0mm 0mm 0mm 3mm},clip]{zvz3.png} 
 \fi
\caption{Young stellar associations  plotted as points on the vertical phase-space
$z_g$ vs $v_z$ distribution for
nearby stars within 200 pc of the Sun using the Gaia DR2 sample.
The vertical axis is the vertical velocity component in km/s. 
The horizontal axis is the height above or below the Galactic plane $z_g$ in pc.
The associations older than 20 Myr are plotted with smaller point sizes.
The middle panel is similar to the top one but only stars and associations with tangential velocity
component $v_\theta > V_{LSR}$ are plotted.
The bottom panel is similar to the top one but only stars and associations with tangential velocity
component $v_\theta < V_{LSR}$ are plotted.
Stellar associations seem to be  associated with peaks in the vertical phase-space distribution.
    \label{fig:zvz}}
\end{figure}

Phase wrapping after a tidal perturbation 
(e.g., \citealt{minchev09,candlish14,delavega15,antoja18,bland19}) can affect
the distribution of stars in phase-space. 
How are the stellar associations distributed in vertical phase-space or as a function 
of $z_g$ and $v_z$?
We plot in Figure \ref{fig:zvz} the current coordinates of the stellar associations
on top of the vertical phase-space distribution of stars in the solar neighborhood.
The ranges in our figure are chosen to encompass the coordinates
of the stellar associations, so our plot only shows stars with orbits that remain 
within 200 pc of the Galactic plane.
The top panel shows the distribution of stars (again from Gaia DR2) within 200 pc 
of the Sun with
the young stellar associations in our sample.  The middle panels only plot stars and
associations that have tangential velocity $v_\theta > V_{LSR}$ and so
shows stars and associations that spend more time at larger galactocentric radius. 
The bottom panel shows stars and associations with $v_\theta < V_{LSR}$.
We have checked 
that the stellar vertical phase-space distributions look similar if the vertical component
of angular momentum is used to choose stars rather than the tangential velocity component.

The morphology of the vertical phase-space distributions in Figure \ref{fig:zvz} shows 
streaks and clumps rather than a spiral that might arise from a simple phase wrapping model
(e.g., \citealt{monari18,antoja18}).  
Prior studies found that the strength of peaks in the solar neighborhood $v_\theta,v_R$
velocity distribution are sensitive to Galactic hemisphere \citep{quillen18_galah,monari18}.  
Here we see that the vertical phase space distribution is sensitive to $v_\theta$.
These two phenomena are probably related. 
This implies that stellar motions
depend on their vertical location or equivalently, as seen here, 
the vertical phase-space distribution is sensitive to angular momentum. 
%$v_\theta$.  These phenomena are probably related. 
%The sensitivity of these peaks  to $v_\theta$ and $v_R$ is probably related to the   sensitivity of peaks in the $v_\theta,v_R$ stellar velocity distribution to Galactic hemisphere  \citep{quillen18_galah,monari18}.
%Figure \ref{fig:zvz} shows that 
%The distribution of stars in nearly planar orbits does not exhibit a spiral-like feature, but neither does it resemble a smooth gaussian distribution.   Clumps in the phase-space distribution are dependent on angular momentum so stars spending more time at larger radius have a different pattern in their vertical motion. 

In Figure \ref{fig:zvz}, 
we have plotted associations younger than 20 Myr with
a larger point size.  This lets us compare a population of objects recently
associated with the distribution and motions in the interstellar medium
to those that are intermediate in age, the rest of the associations  ($>$ 20 Myr but younger than 70 Myr)
and to stars in the solar neighborhood with mean age a few Gyr.
Gas dynamics differs from stellar dynamics as gas can shock, and disturbances in the gas disk will dissipate after a dynamical time.
However, in all three panels in Figure \ref{fig:zvz}, 
the stellar associations seem to be associated with peaks in the vertical phase-space distribution.
This implies that the vertical motions of gas where the stellar associations formed is related 
to the vertical motions of stars that are in nearly planar orbits in the Galactic disk.
Perhaps the gas and stars in nearly planar orbits move vertically together.  

A tidal perturbation on the disk would excite both epicyclic and vertical oscillations.
Even when integrating test particles in a fixed potential, the stellar disk response can be quite complex
\citep{delavega15}.
As stressed by \citet{hunter69,sparke88,darling19}, when a disk bends, 
the potential associated with the perturbation also acts on the unperturbed disc, so 
phase wrapping in a fixed potential (e.g., \citealt{candlish14,delavega15}) does not capture the 
full complexity of the stellar disk response (e.g., \citealt{donghia16}).
The clumps seen in the vertical phase-space distribution could be showing a rippled
disk that was perturbed in the past \citep{quillen09,minchev09,purcell11,chakrabarti11,gomez13,donghia16,bland19,darling19}.
Recent tidal perturbations, such as the interaction of the Antlia 2 dwarf galaxy, on a nearly co-planar orbit  with the Milky Way \citep{chakrabarti19} or the Sagittarius dwarf Galaxy (e.g., \citealt{laporte19}), would also leave visible traces in the gas distribution at present day.  

In the Gaia DR2 sample,  
\citet{antoja18} discovered a spiral in the vertical phase-space distribution of stars in the solar neighborhood  by plotting the distribution of stars as a function of $z_g$ and $v_z$. 
The phase-space spiral \citep{antoja18,bland19,laporte19} 
is present at larger scales in both $v_z$ and $z_g$ than we show
in Figure \ref{fig:zvz}. The ranges in our figure are chosen
to encompass the coordinates
of the stellar associations, however the innermost edge of the spiral 
seen by \citet{antoja18} is at about $|v_z| \sim 20$ km/s and $|z_g| \sim 250$ pc and would
lie outside our plot.   

%In summary, stellar associations are located in between peaks in the $v_\theta, v_R$ stellar velocity distribution for stars in the solar neighborhood.   This is consistent with the hypothesis that the under-densities in the velocity distribution are associated with orbits that are perturbed by spiral density features \citep{quillen18} as the stellar associations were likely to have been born in a spiral arm.
%In contrast, stellar associations seem to be located near peaks in the vertical phase-space distribution ($z_g, v_z$), suggesting that the vertical motions of gas in which stellar associations are born is similar to that of the low velocity dispersion disk stars.

\section{Summary and Discussion}
\label{sec:sum}

In this study we have used recent compilations of membership, space motions, distances and ages
of young (less than  70 Myr old)  clusters, stellar associations, and  star formation regions near the Sun to estimate their 
birth locations.  Our works builds upon efforts of hundreds of prior observational and statistical studies of stars 
(e.g., \citealt{eggen83,delareza89,dezeeuw99,jaya00,mamajek99,binks15,pecaut16,mamajek16,gagne18a}).
 Our backwards orbit integrations are done in a gravitational potential that is an  approximation to the Galactic 
 gravitational potential near the Sun.  This potential is static, separable and axisymmetric and 
 matches recent estimates for the local standard of rest, the rotation curve slope and the 
vertical acceleration as a function of height above the Galactic plane.

Most of the stellar associations were born within the radius of the  Sun and have moved out to the solar
neighborhood where they are now found.  The Octans and Argus associations are exceptions
as they were born at larger galactocentric radius than the Sun.  One way to account
for these trends is with a spiral structure model where nearby spiral features have different
pattern speeds. A spiral arm with pattern speed that is higher than $\Omega_\odot$
places associations  currently on the trailing side of their birth arm.  In this case, N-body simulations 
of flocculent approximately corotating spiral structure \citep{grand12}
and shock models can predict outwards radial motion after birth.
Alternatively  the associations that moved outward were born on the trailing side of a corotating spiral
arm that increased their angular momentum through its gravitational torque  \citep{kawata14}.
%If the local spiral structure is transient and flocculent then most of the young
%associations were born in nearly corotating spiral arms.  
Other scenarios, such as involving tidally excited spiral structure or spurs and armlets 
extending from strong arms might also
account for the radial motions after birth  (e.g., \citealt{dobbs10}).

Variations in  birth heights of the stellar associations suggest that
they were born in a corrugated  disk of molecular clouds, similar to that inferred from
the current filamentary molecular cloud distribution \citep{alves20} and extinction maps \citep{rezaei18,green19,lallement19}.
Maximum vertical heights reached above or below the galactic plane 
are high for the youngest and oldest stellar associations, but only about 30 pc
for the 20--30 Myr older associations such as the $\beta$ Pictoris moving group.    
This implies that there were spatial variations in
the amplitudes of the vertical motions in the  molecular cloud distributions in the recent past.

We examined birth locations in frames rotating at different pattern speeds.  
Multiple arms with  different pattern 
speeds and different heights seem required to account for the stellar association birth locations, suggesting
that spiral arms or molecular cloud filaments  exhibited transient behavior in the past 50 Myr  near the Sun.

We find that  the stellar associations are located in between peaks in the $v_\theta, v_R$ stellar 
velocity distribution for stars in the solar neighborhood.   This supports 
 the hypothesis the dips in the velocity distribution are
 associated with orbits that touch nearby spiral density features \citep{quillen18}.
In contrast, stellar associations seem to be located near peaks in
the  vertical phase-space distribution ($z_g, v_z$), suggesting that  the 
 gas in which stellar associations are born moves together with the 
the low velocity dispersion disk stars.

Ongoing efforts are discovering  new stellar associations (e.g. \citealt{meingast19,gagne20_prep,kounkel19,kounkel20}), 
substructure in known star formation
regions and associations (e.g., \citealt{kos19,tian20}),  improving upon the 
accuracy of membership age distributions, and position and velocity measurements (e.g., \citealt{ujjwal20}).
A backwards integration study can be redone in the future with additional and better measurements and in time dependent and non-axisymmetric potential models.
More distant stellar associations may reveal patterns of star formation in the Local and other arms.
Lastly, improved dissection of N-body simulations that include  gas dynamics and star formation could help 
differentiate between potential spiral arm models and the nature of the vertical motions seen 
in the stars, gas and stellar associations in the solar neighborhood.

%Why do we not see any moving groups coming from the Local arm?  We would if we extended our survey to associations more distance from the Sun.

%Do we agree with the multiple burst hypothesis that has been discussed by \cite{elmegreen93} and \cite{miretroig18}?  Also see \citet{dobbs10}.

%Kinematic age estimates by \citet{crundall19} for the the $\beta$ Pictoris moving group BPMG is $17.8 \pm 1.2$ Myr  and $36.3 \pm 1.4$ Myr for the Tucana-Horologium association (THA).

%\appendix

\begin{table*}  % Tab1
%  \captionsetup{font=large} 
\caption{Stellar associations and moving groups, abbreviations and ages \label{tab:age}}
\begin{tabular}{p{4.8cm}p{1.8cm}p{0.8cm}p{0.7cm}p{5.0cm}}
\hline
Name                  & Abbreviation &  Age   & $\sigma_{\rm age}$  &  Age Reference \\
&& (Myr) & (Myr)  & \\
\hline
$\rho$ Ophiucus  star-forming region   & ROPH   &1&1& \citet{wilking08} \\
Taurus-Auriga star-forming region      & TAU        &1.5&1& \citet{reipurth08} \\
$\epsilon$ Chamaeleontis association & EPSC    &4   &1& \citet{murphy13}  \\
Corona-Australis star-forming region & CRA        &4.5& 0.5& \citet{gennaro10}\\
TW Hydrae  association                      & TWA         &10 & 3 & \citet{bell15} \\
Upper Scorpius  group                       & USCO      &10&3& \citet{pecaut16}\\ %$^*$
Upper Corona-Australis association  & UCRA       &10 &3& \citet{gagne18a} \\ %$^*$
$\eta$ Chamaeleontis cluster          & ETAC         &11& 3 & \citet{bell15} \\
Lower Centaurus Crux group         & LCC            &15&3&\citet{pecaut16} \\ %$^*$
Upper Centaurus Lupus group       & UCL            &16&2& \citet{pecaut16}\\ %$^*$
32 Orionis  group                           & THOR          & 22 & 4 & \citet{bell15}\\
$\beta$ Pictoris moving group    & BPMG          & 24 &3 & \citet{bell15} \\
Octans  association                     & OCT             &35&5& \citet{murphy15} \\
Columba    association                & COL             & 42&5& \citet{bell15}\\
Argus       association                  & ARG             &45&5& \citet{zuckerman19} \\
Carina         association              & CAR             &45 & 8 & \citet{bell15} \\
Tucana-Horologium association & THA             &45 &4  & \citet{bell15}\\
IC2602   cluster                          & IC2602         &46&5& \citet{dobbie10} \\
IC2391   cluster                          & IC2391         &50&5& \citet{barrado04} \\
%Platais 8    cluster                       & PL8            &60&& \citet{platais98} \\
$\mu$ Taurus  association        & MTAU           &62&10&  \citet{gagne20_prep} \\ %$^*$
%Volans-Carina association        & VOLCAR      &  89 & 6  & \citet{gagne20d}\\
%Coma Berenices (CBER), 
%Carina-Near (CARN), 
%the Hyades cluster (HYA), 
%the Pleiades cluster (PLE), 
%the core of the Ursa Major cluster (UMA), 
%AB Doradus (ABDOR),
%$\chi$1 For (XFOR).
% note eta ETAC has age by Mu13 estimated at 4-8
% note ETAC has an eclipsing binary age of 8 +- 0.5 by G10
%118 Taurus group                             & 118TAU     &10 &  &  \\
\hline
 \end{tabular}\\
 %   \captionsetup{font=normalsize} 
{ 
References:
Notes.
Standard deviations in age are estimated from the age range or uncertainty in age in the associated reference. 
\citet{pecaut16} found a larger age spread of about 7 Myr in the Sco-Cen star formation regions.
We adopted the age range of 3 Myr for UCRA and 10 Myr for MTAU as references lacked age range or error estimates.
We adopted the age range for ETAC based on discussion by \citet{bell15}, but also see \citet{gennaro10,murphy13}.
For more discussion on ages and their errors and age distributions see discussions by \citet{riedel17,gagne18a}, 
references therein and the references we have listed here.  
}
\end{table*}

\begin{table*}  %Tab2
\caption{%Positions, Velocities, Extents and Dispersions of Young Stellar Associations 
Central Locations and Variances from the BANYAN Gaussian Models for the Young Stellar Associations
    \label{tab:kin}}
    \centering
    \begin{tabular}{p{1.3cm}p{0.8cm}p{0.8cm}p{0.8cm}p{0.8cm}p{0.7cm}p{0.7cm}p{0.7cm}p{0.6cm}p{0.6cm}p{0.6cm}p{0.6cm}p{0.6cm}p{0.6cm}}
 \hline
Name      & Kin. &$x_h$  & $y_h$ & $z_h$ & $U$  & $V$     & $W$     & $\sigma_x$&$\sigma_y$&$\sigma_z$&$\sigma_U$&$\sigma_V$&$\sigma_W$  \\
                &    ref.   &(pc)&(pc)&(pc)&(km/s)&(km/s)&(km/s)&(pc)&(pc)&(pc)&(km/s)&(km/s)&(km/s)  \\
\hline
ROPH   & G18a  & 124.8& -15.2& 37.6 & -5.9  & -13.5 & -7.9  & 1.33 & 0.51 & 0.66 &1.3 & 4.7& 4.3 \\%&$<2$ & \\
TAU     & G18a   & -116.3 & 6.7 &-35.9 & -14.3 & -9.3  & -8.8 & 11.4 & 10.8 &10.1 & 3.1 & 4.5 & 3.4 \\%&  1-2   & \\
EPSC   & G18a  & 49.9  & -84.8 & -25.6 & -9.9  & -19.3 & -9.7 & 2.5 &  3.6 & 4.0 & 1.6 & 2.2 & 2.0   \\%& 2.3-7  & Mu13  \\
%EPSC   & Bi15   & 54.0  & -92.0 & -26.0 & -10.9& -20.4 & -9.9  &      &        &        &      &        &        \\% & 3-5    & Bi15 \\
CRA     & G18a  & 132.4 & -0.2  & -42.4 & -3.7  & -15.7 & -8.8  &3.71&  0.75&2.04 &1.3 &2.2 & 2.2 \\%& 4-5    & Gen12 \\
TWA     & G18a  & 14.4   & -47.7  & 22.7 & -11.6 & -17.9& -5.6  &12.2&  9.7& 3.9& 1.8 &1.8 & 1.6 \\%& 7-13  & Be15\\
%TWA     & Bi15    & 12.2   & -43.2  & 21.9 & -10.5& -18.3& -5.0 &        &        &      &     &        &    \\% &8 - 12& Bi15 \\
%TWA     & MR18  & 16.0  & -60.0  & 20.0 & -10.9& -20.2 & -5.6 &     &        &        &     &        &       \\%& 3-15 & Ri17 \\
USCO  & G18a  & 121.2 & -17.0 & 48.9 & -4.9  & -14.2 & -6.5 &17.0&  8.2& 8.9& 3.7& 3.2& 2.3   \\%& 7-13 & Pe16 \\
%USCO  &  WM18 & 133.1&-21.3 & 47.6&-6.16 &-16.89&-7.05 & 0.4 & 0.4  & 0.3 & 0.14 & 0.10 &0.09 && \\
UCRA   & G18a & 142.1 & -1.2   & -39.2 &  -3.7 & -17.1&  -8.0&  7.3&  2.4& 5.9& 3.0& 1.8& 1.2   \\%&  10    & \\
%118TAU& G18a & -102.3 & -4.8  & -9.9  & -12.8 & -19.1 & -9.2 & 12.7&  2.4& 1.8& 2.1& 2.8& 1.6  \\%& $\sim 10$ & M16 \\
ETAC    & G18a  & 33.6  & -81.4& -34.8 & -10.0 & -22.3 & -11.7& 0.65& 0.98&0.71&1.6& 2.8& 1.8\\%& 9-14  & Be15 \\
%ETAC    & Bi15   & 33.4  & -81.0 & -34.9 & -10.2 & -20.7 & -11.2&     &        &        &     &        &      \\%& 5-10  & Bi15 \\
%ETAC    & MR18 & 49    & -85    &   -22  & -10.1 & -19.3  & -9.7 &     &        &        &     &        &      \\%& 5-8    & Ri17 \\
LCC     & G18a   & 54.3 &-94.2  & 5.8    & -7.8   & -21.5  & -6.2 &11.9& 12.4&13.7& 2.7& 3.8& 1.8 \\%&  12-18   & Pe16 \\
UCL     & G18a   &107.5& -60.9 & 26.5  & -4.7   & -19.7  & -5.2  &21.0&19.6&13.5& 3.8& 3.0& 1.7 \\% & 14-18 & Pe16\\
%UCL     & WM18 &114.0 & -65.5 & 34.6 -& -5.90 & -20.00 &-5.80&0.5 & 
THOR  & G18a   & -88.4 & 25.7 & -23.9 & -12.8 & -18.8 & -9.0  &4.1  &  6.9 & 5.1& 2.2 & 2.2 & 2.0 \\%& 19-26 & \\
BPMG  & G18a   & 4.1   & -6.7  & -15.7 & -10.9  & -16.0  & -9.0 &29.3& 14.0& 9.0& 2.2 & 1.2& 1.0\\%& 24     & Be15\\
%BPMG  & Bi15    &  4.3  &  -5.8 & -13.3 &  -11.2 & -16.2 & -9.3   &      &        &       &     &        &      \\%& 21-26  & Bi15 \\
%BPMG  & MR18  & 17.0 & -4.0  & -15.0 & -9.9   & -16.2 & -9.0   &      &        &       &     &        &      \\%& 10-24  & Ri17\\
OCT     & G18a  & 4.0    & -96.9 & -59.7& -13.7 & -3.3   & -10.1 &78.3& 25.8& 8.8& 2.4 & 1.3&1.4 \\%& 30-40 & Mu15\\
%OCT    &  MR18 & -9.0   &-100.0& -61.0& -13.9 & -3.6   & -10.3  &      &       &       &        &      &      \\%& 20-40 & Ri17\\
COL    & G18a   & -25.9  & -25.9& -21.4 & -11.9 & -21.3 & -5.7  &12.1&23.0 &17.8& 1.04&1.29&0.75\\%& 38 -48  & Be15\\
%COL    & MR18  & -29.0  & -48.0 & -37.0& -12.8 & -21.8 & -5.5  &       &        &       &        &       &      \\% & 30-42 & Ri17 \\
ARG    & Z19/G20& 4.9  & -43.3 & -7.8  & -22.8 & 14.1  & -5.0  & 28.9& 41.3&19.2& 1.2 &2.0 & 1.7  \\%&  40-50  & Z19   \\
%ARG    & MR18 & 12.0   &-120.0& -16.0 & -22.6 & -14.1 & -5.3 &      &        &       &        &       &       \\%&  35-50   & Ri17  \\
%ARG    & Bi15   & 14.6   & -24.7& -6.7   & -21.8 & -12.1 & -4.5 &       &        &       &        &       &       \\%&   30-50 & Bi15 \\
CAR    & G18a  & 6.7    & -50.5 & -15.5 & -10.7 & -21.9 & -5.5 & 10.0&18.1&12.6& 0.67&1.02&1.01\\%&38-56   & Be15 \\
%CAR    & Bi15   & 15.5  & -58.5 & -23.0  & -10.5 & -22.4 & -5.8 &      &        &       &        &       &       \\%&  20-40 & Bi15 \\
%CAR    & MR18 & 8.0   & -94.0  & -17.0 & -10.3 & -22.8  & -4.6 &      &        &       &        &       &       \\%&  30-45  & Ri17  \\
THA    & G18a  &  5.4  & -20.1  & -36.1 & -9.8    & -20.9  & -1.0 &19.4& 12.4& 3.8& 0.87&0.79&0.72\\%& 41-49 & Be15 \\
%THA    & MR18 & 0.0   & -22.0  & -35.0 & -9.8   & -20.9 & -1.1   &      &        &       &        &       &       \\%&  30-45 & Ri17 \\
%THA    & Bi15   & 11.8  & -20.8  & -35.7 &-9.9    &-20.7  & -0.9   &      &        &       &        &       &      \\% &  39-43 & Bi15 \\
IC2602& G18a & 47.4 & -137.6 & -12.6 & -8.2   & -20.6 & -0.6  &1.5 & 5.4 & 1.1 & 1.18 &2.61 &0.65\\%&41-52  & Do10\\
IC2391& G18a & 1.9   & -148.1 & -18.0 &-23.0 & -14.9  & -5.5  &1.3&  6.4 & 1.4 & 1.10&3.40 &0.78 \\%& 45-55  & Ba04 \\
%PL8      & G18a & 10.6 & -124.5& -13.9 & -11.0 & -22.9 & -3.6  &7.0& 11.6& 4.5& 1.15&1.96 &0.74\\%& $\sim 60$ & \\
MTAU &  G20   &-130.7&  0.2   & -79.7 & -14.2 & -24.2 &  -6.2 &21.9& 20.8& 12.4& 3.0& 1.7& 2.4 \\%& 62       &   G20 \\
%VOLCAR &G18d & 21.6 & -80.3 & -14.3& -16.1& -28.1&  -0.8  &  3.9&  5.0 & 4.5& 1.4 & 1.4& 1.0 \\  %89,   6,
\hline
%PLE     & G18a    & -118.9 & 28.5   & -54.4  & -6.7   & -28      & -14.0  &112 & \\
%CARN    & G18a    & 0.7      & -28.1  & -4.3   & -25.3  & -18.1  & -2.3 & 200 & 3\\
%CBER    & G18a    & -6.0     & -5.1    & 84.9   & -2.3   & -5.5  & -0.6 & 562 & \\
% HYA      & G18a   & -38.5   & 0.8     & -15.8  & -42.3 & -18.8  & -1.5   &  750     &\\
%ABDOR   & G18a   & -6       & -7.2   & -8.8   & -7.2    & -27.6   & -14.2 &  149 &  Be15\\
%ABDOR   &  Bi15    & -2.2    & 2.9     & -15.4 & -7.1   & -27.2  & -13.8 & 70-120 & Bi15 \\
%ABDOR  &  MR18    & 5.0    & -14.0   & -20.0  & -6.7   & -27.5  & -14.0 & 50-150 & Ri17 \\\hline
%  UMA     & G18a  & -7.5     & 9.9      & 21.9   & 14.8   & 1.8     & -10.2 & 414 &\\
%XFOR    & G18a  & -27.1   & -46.3  & -84.2  & -12.5 & -22.2  & -6.3   & 500 & \\\hline
% Orion Complex clusters     & & & & & & & \\\hline
%  ASCC 16 & K19  & -310.9 & -121.7 & -111.9  & -19.7  & -8.4  & 2.6  &   &  \\
%  ASCC 18 & K19  & -363.3 & -150.2 & -132.7  & -27.2 & -8.4  & -0.1  &&\\
%  ASCC 20 & K19  & -329.9  & -129.9 & -111.7 & -27.7  & -8.2  & -1.3  && \\
%  ASCC 21 & K19  & -313.9  & -113.4 & -100.2 & -19.1  & -8.2 & 3.2   &&   \\
% ASCC 21a& K19  & -318.6  & -125.0 & -101.9 & -19.5  & -9.2 & 3.6   &&\\\hline
% the error in what is in table 1 by MR18 and xyz is only the y, here x,z seem the same.
    \end{tabular}\\
%    \captionsetup{font=normalsize} 
%\caption%This table includes all the objects from Gagne's, Miret-Roig, and Kos' paper with their initial 6D kinematics (XYZUVW). Gagne's data is unchanged, but we had convert the 6D proper motion values into XYZUVW and also the curvlinear coordinate into XYZUVW.  (Note: For Gagne's and Kos's data, the solar radius $R\textsubscript{\(\odot\)}$ used to calculate the values in these table are $R\textsubscript{\(\odot\)} = 8 kpc$ and the circular tangential velocity for the sun $v\textsubscript{\(\odot\)} = 220 km/s$, where as for Miret-Roig's data, $R\textsubscript{\(\odot\)} = 8.5 kpc$ but the circular tangential velocity for the sun remains the same. %G18d =\citet{gagne18d};
{References:  G18a =\citet{gagne18a}; Z19  = \citet{zuckerman19};  
G20 = \citet{gagne20_prep} %{\color{red}[ARP: are these ordered in any particular way? If not, ordering by ref. or Name may be useful.]}.
}
 \end{table*}

\begin{table*} %Tab3
\caption{
Birth sites and other orbital parameters for Young Stellar Associations
    \label{tab:birth}}
    \centering
\begin{tabular}{lrrrrrrrrr}
\hline
name    & $z_b$ & $v_{z,b}$ & $R_b$ & $v_{R,b}$   &$\theta_b-\theta_\odot$& $R_L$ & $|z|_{max}$ & eccentricity  \\
        & (pc)   & (km/s) & (kpc)  & (km/s)   & (rad)            & (kpc)        &  (pc) & \\
\hline
ROPH    & $ 44\pm  3$ & $-0.5\pm 3.4$ & $8.002\pm 0.004$ & $-5.1\pm 1.0$ & $0.031\pm 0.021$ & $7.962\pm 0.185$ & $ 56\pm 41$ & $ 0.0228\pm  0.0070$ & \\
TAU     & $-28\pm 13$ & $-1.7\pm 3.6$ & $8.234\pm 0.012$ & $ 2.9\pm 3.7$ & $0.042\pm 0.023$ & $8.351\pm 0.124$ & $ 52\pm 48$ & $ 0.0281\pm  0.0072$ & \\
EPSC    & $-10\pm  9$ & $-2.7\pm 1.9$ & $8.074\pm 0.005$ & $ 0.4\pm 1.1$ & $0.123\pm 0.026$ & $7.849\pm 0.063$ & $ 59\pm 23$ & $ 0.0294\pm  0.0086$ & \\
CRA     & $-29\pm 11$ & $-2.2\pm 2.3$ & $8.021\pm 0.008$ & $-6.4\pm 1.4$ & $0.129\pm 0.014$ & $7.883\pm 0.070$ & $ 41\pm  7$ & $ 0.0331\pm  0.0075$ & \\
TWA     & $  7\pm 16$ & $ 2.4\pm 1.1$ & $8.087\pm 0.022$ & $ 3.6\pm 2.1$ & $0.287\pm 0.080$ & $7.913\pm 0.061$ & $ 42\pm 19$ & $ 0.0239\pm  0.0038$ & \\
USCO    & $ 37\pm 24$ & $ 2.7\pm 2.1$ & $8.056\pm 0.037$ & $-4.6\pm 3.8$ & $0.289\pm 0.081$ & $7.927\pm 0.112$ & $ 67\pm 11$ & $ 0.0222\pm  0.0116$ & \\
UCRA    & $-20\pm 11$ & $-1.9\pm 0.8$ & $8.038\pm 0.032$ & $-4.1\pm 3.2$ & $0.284\pm 0.080$ & $7.803\pm 0.055$ & $ 38\pm  7$ & $ 0.0351\pm  0.0078$ & \\
ETAC    & $ 23\pm 26$ & $-4.6\pm 1.7$ & $8.066\pm 0.021$ & $ 5.1\pm 2.4$ & $0.314\pm 0.078$ & $7.736\pm 0.078$ & $ 89\pm 25$ & $ 0.0409\pm  0.0130$ & \\
LCC     & $ -7\pm 26$ & $ 1.2\pm 1.2$ & $8.057\pm 0.048$ & $ 4.7\pm 3.9$ & $0.427\pm 0.079$ & $7.730\pm 0.109$ & $ 37\pm 16$ & $ 0.0462\pm  0.0099$ & \\
UCL     & $-11\pm 26$ & $ 2.8\pm 1.2$ & $8.057\pm 0.061$ & $ 1.2\pm 3.9$ & $0.455\pm 0.053$ & $7.749\pm 0.084$ & $ 56\pm 14$ & $ 0.0379\pm  0.0113$ & \\
THOR    & $ 24\pm 33$ & $-1.4\pm 0.7$ & $8.094\pm 0.060$ & $ 8.3\pm 2.7$ & $0.618\pm 0.107$ & $8.014\pm 0.072$ & $ 53\pm 20$ & $ 0.0296\pm  0.0108$ & \\
BPMG    & $ 27\pm 13$ & $-0.7\pm 0.6$ & $8.065\pm 0.069$ & $ 4.3\pm 2.4$ & $0.683\pm 0.081$ & $7.984\pm 0.063$ & $ 28\pm 13$ & $ 0.0171\pm  0.0035$ & \\
OCT     & $ 69\pm 18$ & $-1.0\pm 1.5$ & $8.323\pm 0.127$ & $-12.1\pm 1.9$ & $1.041\pm 0.135$ & $8.438\pm 0.104$ & $ 69\pm 17$ & $ 0.0361\pm  0.0056$ & \\
COL     & $  4\pm 17$ & $-1.9\pm 0.8$ & $7.773\pm 0.076$ & $12.6\pm 1.7$ & $1.201\pm 0.142$ & $7.845\pm 0.040$ & $ 24\pm  6$ & $ 0.0424\pm  0.0058$ & \\
ARG     & $ -6\pm 19$ & $-2.2\pm 1.2$ & $9.026\pm 0.211$ & $-38.4\pm 3.3$ & $1.364\pm 0.123$ & $9.034\pm 0.093$ & $ 63\pm 28$ & $ 0.1144\pm  0.0090$ & \\
CAR     & $  3\pm 18$ & $-1.9\pm 1.1$ & $7.704\pm 0.106$ & $13.4\pm 1.9$ & $1.293\pm 0.229$ & $7.789\pm 0.036$ & $ 31\pm 18$ & $ 0.0440\pm  0.0033$ & \\
THA     & $ -1\pm 25$ & $-6.6\pm 0.8$ & $7.770\pm 0.059$ & $12.3\pm 1.2$ & $1.285\pm 0.114$ & $7.809\pm 0.030$ & $100\pm 12$ & $ 0.0396\pm  0.0045$ & \\
IC2602  & $-19\pm 30$ & $-6.6\pm 0.8$ & $7.767\pm 0.131$ & $12.2\pm 3.5$ & $1.330\pm 0.142$ & $7.784\pm 0.087$ & $105\pm  9$ & $ 0.0384\pm  0.0112$ & \\
IC2391  & $ 15\pm  8$ & $-1.7\pm 0.8$ & $7.742\pm 0.182$ & $-3.1\pm 4.7$ & $1.534\pm 0.148$ & $8.055\pm 0.121$ & $ 32\pm  9$ & $ 0.0405\pm  0.0037$ & \\
MTAU    & $ 64\pm 33$ & $ 2.6\pm 2.5$ & $7.461\pm 0.124$ & $ 6.3\pm 5.5$ & $1.828\pm 0.217$ & $7.826\pm 0.066$ & $ 81\pm 14$ & $ 0.0509\pm  0.0084$ & \\
\hline

  \end{tabular}\\
{}
 \end{table*}

\vskip 0.2 truein

\section*{Acknowledgements}

This project was developed in part at the 2019 Santa Barbara Gaia Sprint, hosted by the Kavli Institute for Theoretical Physics (KITP) at the University of California, Santa Barbara.
This research was supported in part at KITP by the Heising-Simons Foundation and the US National Science Foundation under Grant No. NSF PHY-1748958.  SC acknowledges support from NASA ATP NNX17AK90G,
NSF AAG grant 1517488, and from Research Corporation
for Scientific Advancement's Time Domain Astrophysics
Scialog.
We thank Eric Mamajek, Borja Anguiano, Dechen Dolker, and Tim Crundall for helpful discussions and correspondence.

\section*{Data Availability}

New data generated are incorporated into the article.

\bibliographystyle{mnras}
\bibliography{refs_gal} 

\end{document}